%% file: NSB_let_revised.tex
\documentclass[aps,prl,reprint,superscriptaddress,twocolumn,hidelinks,nofootinbib]{revtex4-2}

\usepackage{graphicx}

\usepackage{rotating}

\usepackage{dcolumn}

\usepackage{amssymb}

\usepackage{amsmath}

\usepackage{bm}

\usepackage{listings}

\usepackage{fancyhdr}

\usepackage{wrapfig}

\usepackage{subfig}

\usepackage{xcolor}

\usepackage{hyperref}

\hypersetup{
    colorlinks,
    linkcolor={red!50!black},
    citecolor={blue!50!black},
    urlcolor={blue!80!black}
}

\include{mydefs}

\allowdisplaybreaks

\begin{document}

\title{The spatial gauge-dependence of single-field inflationary bispectra}

\author{Ermis Mitsou}
\email[]{ermitsou@physik.uzh.ch}
\affiliation{Center for Theoretical Astrophysics and Cosmology, Institute for Computational Science, University of Zurich, CH--8057 Zurich, Switzerland}

\author{Jaiyul Yoo}
\email[]{jyoo@physik.uzh.ch}
\affiliation{Center for Theoretical Astrophysics and Cosmology, Institute for Computational Science, University of Zurich, CH--8057 Zurich, Switzerland}
\affiliation{Physics Institute, University of Zurich, Winterthurerstrasse 190, CH--8057, Zurich, Switzerland}

\date{\today}

\begin{abstract}
In single-field inflationary models the bispectra are usually given in the $\ze$-gauge, because its temporal part leads to the super-horizon conservation of fluctuations. However, this property is independent of the choice of {\it spatial} gauge, so in this letter we explore this freedom. We compute the variation of the bispectra under the most general spatial gauge transformation that is globally defined and privileges no point, direction or scale. In the squeezed configuration we then obtain a generalization of the classic $\ze$-gauge consistency relations, which we also derive through the `large diffeomorphism' approach for all four bispectra. At leading order in the long wave-number the transformation only affects the case where the long mode is a scalar. The first effect is a shift of the tilt factor, so that one can significantly reduce the amplitude of that contribution. Secondly, there is now an extra term depending on the triangle shape, the same as in solid inflation, which is due to the fact that the 3-metric has a scalar anisotropy in generic spatial gauge. At next-to-leading order there is no variation, so the conformal consistency relations of the $\ze$-gauge are preserved.
\end{abstract}

\maketitle

Understanding the early universe, and in particular its deviation from Gaussian primordial fluctuations, is one of the major goals of the upcoming cosmological surveys \cite{SKA,EUCLID,WFIRST,DESI,LSST}. The simplest measure of non-Gaussianity on a time-hypersurface is the 3-point correlation function, or, in Fourier space, the bispectrum 
\bea
\bra X(t,\bm{k}_1)\, Y(t,\bm{k}_2)\, Z(t,\bm{k}_3) \ket & = & (2\pi)^3 \de^{(3)} \( \bm{k}_1 + \bm{k}_2 + \bm{k}_3 \) \nn \\
 & & \times B_{XYZ}(t,k_1,k_2,k_3) \, , \label{eq:Bdef}
\eea
where statistical homogeneity and isotropy are assumed. At the level of observables, the bispectrum of primordial fluctuations \eqref{eq:Bdef} enters the expressions of the 3-point statistics in the sky (along with light-propagation effects), but its squeezed limit $k_3 \ll k_1 \approx k_2$ also affects the 2-point statistics of galaxies through the bias model \cite{Dalal:2007cu,Matarrese:2008nc}. Moreover, in the case of single-field inflationary models, the squeezed bispectrum satisfies a particular `consistency relation' to the power spectrum \cite{Maldacena:2002vr,Creminelli:2004yq,Cheung:2007sv,Ganc:2010ff,Creminelli:2011rh,Pimentel:2013gza}\footnote{With some notable caveats \cite{Kinney:2005vj,Chen:2010xka,Namjoo:2012aa,Martin:2012pe,Chen:2013aj,Mooij:2015yka,Suyama:2021adn,Avis:2019eav} that lead to a different or more general consistency relation \cite{Bravo:2017wyw,Finelli:2017fml,Finelli:2018upr,Bravo:2020hde}.}
\beq \label{eq:Pdef}
\bra X(t,\bm{k}_1)\, Y(t,\bm{k}_2) \ket = (2\pi)^3 \de^{(3)} \( \bm{k}_1 + \bm{k}_2 \) P_{XY}(t,k_1) \, ,
\eeq
which is an instance of more general inflationary $N$-point function `soft-theorems' \cite{Chen:2006dfn,Creminelli:2012ed,Senatore:2012wy,Hinterbichler:2012nm,Assassi:2012zq,Hinterbichler:2013dpa,Goldberger:2013rsa,Kundu:2014gxa,Kundu:2015xta,Shukla:2016bnu,Gong:2017wgx,Pajer:2017hmb,Bordin:2017ozj,Hui:2018cag,Jazayeri:2019nbi}. The corresponding relation in the observable statistics is then a falsifiable prediction, which would therefore rule out this vast class of models if not confirmed by future surveys.   

Importantly, the primordial bispectrum \eqref{eq:Bdef} itself is not an observable quantity and, unlike observables, it is coordinate-dependent. This freedom can therefore be exploited to simplify the theoretical description of the physics under consideration, both computationally and conceptually. For the temporal coordinate ambiguity, i.e. the choice of time-hypersurface foliation, single-field models offer a physically privileged option, the `unitary gauge', where the hypersurfaces are those of constant inflaton value. It has the computational advantage of leading to fluctuations that are conserved \cite{Salopek:1990jq,Wands:2000dp,Maldacena:2002vr,Weinberg:2003sw,Lyth:2004gb,Weinberg:2008zzc,Naruko:2011zk,Senatore:2012ya} and classical \cite{Starobinsky:1982ee,Lyth:2006qz} on super-horizon scales, so that the spectra freeze after the modes exit the horizon. This is crucial for having an unambiguous `output' from the inflationary era which can then turn into a continuous `input' to the big-bang phase, as the modes re-enter the horizon, independently of the unknown intermediate sub-horizon physics. We are thus left with the freedom of the spatial parametrization of those hypersurfaces, as super-horizon conservation is independent of that choice \cite{Lyth:2004gb}. 

One known effect of spatial coordinate transformations is that they can actually eliminate the squeezed primordial bispectrum altogether in single-field models \cite{Baldauf:2011bh,Tanaka:2011aj,Creminelli:2012ed,Senatore:2012ya,Pimentel:2012tw,Senatore:2012wy,Pajer:2013ana,dePutter:2015vga,Dai:2015jaa,Dai:2015rda,Cabass:2016cgp,Bravo:2017gct,Cabass:2018roz,Umeh:2019qyd}. This feature significantly simplifies many computations, but perhaps more importantly, it confirms the physical intuition that large and small scales should decouple, up to coordinate artifacts, at least within a finite observation volume. All known coordinate systems with this property are essentially different versions of the same principle -- eliminating the long-wavelength metric fluctuations in the vicinity of the observer world-line -- with the most developed and established representative being the `conformal Fermi coordinates' (CFC) \cite{Pajer:2013ana,Dai:2015rda} (see \cite{Mitsou:2020czr} for a light-cone-adapted version of the idea). 

One disadvantage of such coordinates is the fact that they are defined through a series expansion in the spatial coordinates, so that they privilege a point in space and one can only use them in a spatial patch of finite extent in practice. Although the CFC were precisely conceived so that this patch is large enough for the purpose of relating inflation to observations, this type of coordinate system is qualitatively different, and less convenient, than the usual choices in cosmological perturbation theory. These are the so-called `gauges', in the sense of Bardeen \cite{Bardeen:1980kt}, which are defined globally in space and privilege no point, direction, or scale. On top of being simpler to work with computationally, they also allow one to use the standard tools of cosmological perturbation theory straightforwardly and unambiguously, i.e. the Fourier transform, scalar-vector-tensor decomposition, gauge transformations, etc. In contrast, the use of a series expansion in real space, as in CFC, unavoidably obscures the global behavior of the fields and thus complicates the use of the aforementioned machinery. However, apart from the CFC-like systems, alternative spatial gauges have not been explored in the literature to our knowledge. Here we therefore compute the variation of the bispectra of single-field models under the most general spatial gauge transformation that is globally defined and privileges no point, direction or scale. 

The most noteworthy result is the corresponding generalization of the consistency relation, which is altered in the case where the long mode is a scalar. Let us first remind the classic result in the $\ze$-gauge \cite{Maldacena:2002vr,Creminelli:2004yq,Cheung:2007sv}
\beq \label{eq:zegres}
B_{\ze_L X_S X_S} \approx - \, P_{\ze*}(k_L) \, P_{X*}(k_S) \, n_X(k_S) \, ,
\eeq 
where $\ze$ is the curvature mode, $X$ is either $\ze$ or the tensor modes $\ga_a$, $k_L \ll k_S$ are the long and short mode wave-numbers, respectively, the asterisk denotes evaluation at the respective horizon-crossing times $t = t_*(k)$ and
\beq
n_X(k) := \frac{\ed \log \[ k^3 P_{X*}(k) \]}{\ed \log k} \, , 
\eeq
is the power spectrum tilt (conventionally $n_s-1$ for the scalar). Note that here we focus on the squeezed {\it limit}, i.e. only the leading order terms in $k_L$ -- the subleading ones will be discussed at the end. Now, the transformed squeezed bispectrum depends on a single function $\be(t)$
\bea
B_{\ti{\ze}_L\ti{X}_S\ti{X}_S} & \approx & - \, P_{\ti{\ze}*}(k_L) \, P_{\ti{X}*}(k_S) \[ \frac{}{} n_{\ti{X}}(k_S) + \be_*(k_L) \right. \label{eq:Bresult} \\
 & & \hspace{1cm} \left. -\, \be_*(k_L) \( 3 - n_{\ti{X}}(k_S) \) (\hat{\bm{k}}_S \cdot \hat{\bm{k}}_L)^2 \frac{}{} \] \, ,  \nn
\eea
where the tilde denotes the fields in the new gauge. In the first line we see that we have shifted the tilt factor, although with a function of $k_L$ instead of $k_S$. In the second line we see that we have also generated a quadrupole term $\sim ( \hat{\bm{k}}_S \cdot \hat{\bm{k}}_L )^2$, which introduces a dependence on the shape of the momentum configuration with which we take the squeezed limit. The reason for this is that $\be$ controls the anisotropic scalar component of the metric $\sim \pa_i \pa_j$, which therefore introduces a dependence on momentum direction. This is a similar situation to solid inflation, where the squeezed bispectrum also has a quadrupolar component, because the scalar mode induces an anisotropic deformation \cite{Endlich:2012pz,Endlich:2013jia}. 

Focusing on the case of isosceles triangles $\hat{\bm{k}}_S \cdot \hat{\bm{k}}_L = 0$, or angle-averaged observables, so that only the first line of \eqref{eq:Bresult} survives, we note that the choice
\beq \label{eq:be}
\be_*(k) = - \, n_X(k) \, ,
\eeq
leads to (up to $\Ord(k_L/k_S)$ corrections, see last section)
\beq \label{eq:Brun}
B_{\ti{\ze}_L \ti{X}_S \ti{X}_S} \os{\rm iso./ang.-av.}{\approx} - \, P_{\ti{\ze}*}(k_L) \, P_{\ti{X}*}(k_S) \, \al_{\ti{X}}(k_S) \, , 
\eeq
where
\beq
\al_X(k) := \frac{\ed \, n_X(k)}{\ed \log k} \, , 
\eeq
is the running of the power spectrum. This quantity is generically suppressed by one order of slow-variation parameters compared to the tilt \cite{Maldacena:2002vr,Chen:2006nt,Bartolo:2010im}, so it is negligible, at least for viable models, given the observational constraint of nearly scale-invariant spectrum. In practice the tilt is given by background functions of time, such as the slow-variation parameters and speed of sound, evaluated at horizon-crossing time. One can therefore choose $\be$ to be equal to that combination of time-functions in order to reproduce \eqref{eq:be}. For instance, in slow-roll inflation the scalar case $\ti{X} = \ti{\ze}$ would be given by the combination of the first two slow-variation parameters $\be = 2 \vep + \et$.

{\it Spatial gauge transformation} -- We start with the exponential parametrization of the 3-metric \cite{Maldacena:2002vr}
\beq \label{eq:pertdef}
g_{ij} = a^2 \[ e^{2 \bm{h}} \]_{ij} \equiv a^2 \[ \de_{ij} + 2 h_{ij} + 2 h_{ik} h_{kj} + \dots \] \, ,
\eeq
where $a$ is the scale factor and the Latin indices are displaced/contracted with $\de_{ij}$. Note that we need to work to second order in the perturbations, since the linear-order bispectrum is trivially zero, hence the relevance of specifying the relation \eqref{eq:pertdef} non-linearly. Following standard practice \cite{Maldacena:2002vr}, the $g_{0\mu}$ components of the metric can be integrated out of the action to also become functionals of $h_{ij}$ and the inflaton field. The standard $\ze$-gauge is given by zero inflaton fluctuations (unitary temporal gauge) and
\beq \label{eq:hijze}
h_{ij} = \de_{ij} \ze + \ga_{ij} \, , \hspace{1cm} \pa_i \ga_{ij}, \ga_{ii} \equiv 0 \, .
\eeq
We now want to transform this to the most general spatial gauge that is globally defined and privileges no point, direction, or scale. Given the absence of vector degrees of freedom, this is given by
\beq \label{eq:hijnew}
h_{ij} = \( \de_{ij} + \be \pa^{-2} \pa_i \pa_j \) \ti{\ze} + \ti{\ga}_{ij} \, , \hspace{0.7cm} \pa_i \ti{\ga}_{ij}, \ti{\ga}_{ii} \equiv 0 \, ,
\eeq
where $\ti{\ze}$ and $\ti{\ga}_{ij}$ are the scalar and tensor fields in that gauge and $\be$ is a dimensionless function of time. Note that we could also allow $\be$ to depend on $\pa^2/\cH^2$, where $\cH$ is the conformal Hubble parameter, but this quantity is zero at super-horizon scales. The two gauges are related by a spatial gauge transformation of the form 
\beq \label{eq:trans}
x^i \to x^i - \be \pa^{-2} \pa_i \ze + \Ord(2) \, ,
\eeq
where the second-order part is chosen such that \eqref{eq:hijnew} holds up to second order and is irrelevant for our purposes here. Indeed, we only need \eqref{eq:pertdef}, \eqref{eq:hijze}, \eqref{eq:hijnew} and the first-order part of \eqref{eq:trans} to find the map between the fields to second order  
\bea
\ti{\ze} & = & \[ 1 + \be \( \pa^{-2} \pa_i \ze \) \pa_i \] \ze + \frac{1}{2}\, P_{ij} \xi_{ij} + \Ord(3)  \, , \label{eq:zeti} \\
\ti{\ga}_{ij} & = & \ga_{ij} + \( P_i^k P_j^l - \frac{1}{2}\, P_{ij} P^{kl} \) \xi_{kl} + \Ord(3) \, , \label{eq:gati}
\eea
with $P_{ij} := \de_{ij} - \pa^{-2} \pa_i \pa_j$ the transverse projector and 
\beq
\xi_{ij} := \be \( \pa^{-2} \pa_k \ze \) \pa_k \ga_{ij} - \frac{1}{2}\, \be^2 \( \pa^{-2} \pa_i \pa_k \ze \) \( \pa^{-2} \pa_j \pa_k \ze \) \, .
\eeq
We now work for definiteness in \eqref{eq:hijnew} and in Fourier space
\beq \label{eq:gauge}
h_{ij} = \( \de_{ij} + \be \hat{k}_i \hat{k}_j \) \ti{\ze} + \ti{\ga}_a e^a_{ij} \, ,
\eeq
where $\{ e_{ij}^a(\hat{\bm{k}}) \}_{a=1,2}$ is a basis of traceless-transverse polarization tensors
\beq
e_{ii}^a \equiv 0 \, , \hspace{1cm} k^j e_{ij}^a(\hat{\bm{k}}) \equiv 0 \, , \hspace{1cm} e_{ij}^a e_{ij}^b \equiv \de^{ab} \, .
\eeq
The field relations \eqref{eq:zeti} and \eqref{eq:gati}, which take the form $\ti{X} = X + X^{(2)}$, provide the bispectrum relation \cite{Maldacena:2002vr} 
\bea
\bra \ti{X}_1 \ti{Y}_2 \ti{Z}_3 \ket & = & \bra X_1 Y_2 Z_3 \ket + \bra X^{(2)}_1 Y_2 Z_3 \ket  \label{eq:bisrel} \\
 & & +\, \bra X_1 Y^{(2)}_2 Z_3 \ket + \bra X_1 Y_2 Z^{(2)}_3 \ket + \Ord(5) \, , \nn
\eea
where we used the notation $X_n := X(\bm{k}_n)$. The 4-point statistics are then dominated by their free 2-point contributions, up to trispectrum and other higher-order corrections. For instance, for a field product $V W$ inside the real-space $Z^{(2)}$, we find a convolution in Fourier space
\beq
\bra X_1 Y_2 (V * W)_3 \ket \propto P_{XV}(k_1) \, P_{YW}(k_2) + V \leftrightarrow W \, ,
\eeq
up to the overall Dirac delta. In the present case, the non-trivial free power spectra are
\beq \label{eq:Pkfree}
P_{\ze\ze} = P_{\ze} \, , \hspace{1cm} P_{\ga_a\ga_b} = P_{\ga} \de_{ab} \, ,
\eeq 
and they are the same for both tilded and untilded fields, since these are equal linearly (ditto for $n_X, \al_X$). To compute \eqref{eq:bisrel} we use the momentum conservation $\bm{k}_1 + \bm{k}_2 + \bm{k}_3 = 0$ to eliminate all scalar products, e.g. $\bm{k}_1 \cdot \bm{k}_2 = \( k_3^2 - k_1^2 - k_2^2 \) / 2$. The difference between the two gauges $\De B_{XYZ} := B_{\ti{X}\ti{Y}\ti{Z}} - B_{XYZ}$ is then given by
\begin{widetext}
\bea
\De B_{\ze_1\ze_2\ze_3} & = & - \,\be P_{\ze,2} P_{\ze,3} \[ 1 + \frac{1}{2} \( \frac{k_2^2 - k_1^2}{k_3^2} + \frac{k_3^2 - k_1^2}{k_2^2} \) \right. \\
 & & \left. +\, \be \, \frac{\( k_1 + k_2 + k_3 \) \( k_1 - k_2 + k_3 \) \( k_1 + k_2 - k_3 \) \( k_1 - k_2 - k_3 \) \( k_1^2 - k_2^2 - k_3^2 \)}{8 \( k_1 k_2 k_3 \)^2} \] + {\rm cyc.} \, , \label{eq:DeBsss}  \\
\De B_{\ze_1\ze_2\ga_{a,3}} & = & - \frac{1}{4}\, \be \hat{k}_1^i \hat{k}_2^j e_{a,3}^{ij} \[ \( P_{\ze,1} - P_{\ze,2} \) P_{\ga,3} \( \frac{k_1}{k_2} - \frac{k_2}{k_1} \) + \( P_{\ze,1} + P_{\ze,2} \) P_{\ga,3} \, \frac{k_3^2}{k_1 k_2}  \right. \nn \\
 & & \hspace{2.4cm} \left. +\, 2 \be P_{\ze,1} P_{\ze,2} \( \frac{k_3^2}{k_1 k_2} - \frac{k_1}{k_2} - \frac{k_2}{k_1} \) \] \, , \\
\De B_{\ze_1\ga_{a,2}\ga_{b,3}} & = & - \frac{1}{2}\, \be e_{a,2}^{ij} e_{b,3}^{ij} P_{\ze,1} \[ P_{\ga,2} + P_{\ga,3} + \( P_{\ga,2} - P_{\ga,3} \) \frac{k_2^2 - k_3^2}{k_1^2} \] \, , \\
\De B_{\ga_{a,1}\ga_{b,2}\ga_{c,3}} & = & 0  \, ,  \label{eq:DeBttt}
\eea
and the non-trivial squeezed limits are
\bea
\De B_{\ze_L\ze_1\ze_2} & = & - \, \be P_{\ze,L} P_{\ze,S} \[ 1 - \( 3 - n_{\ze}(k_S) \) (\hat{\bm{k}}_S \cdot \hat{\bm{k}}_L)^2 + \Ord \( \frac{k_L^2}{k_S^2} \) \] \, , \label{eq:DeBsssSq} \\
\De B_{\ze_L\ga_{a,1}\ga_{b,2}} & = & - \, \be P_{\ze,L} P_{\ga,S} \[ 1 - \( 3 - n_{\ze}(k_S) \) (\hat{\bm{k}}_S \cdot \hat{\bm{k}}_L)^2 + \Ord \( \frac{k_L^2}{k_S^2} \) \] \de_{ab} \, , \label{eq:DeBsttSq} \\
\De B_{\ga_{a,L}\ze_1\ze_2} & = & - \, \be P_{\ga,L} P_{\ze,S} \times \Ord \( \frac{k_L^2}{k_S^2} \) \, , \label{eq:DeBtssSq}
\eea
\end{widetext}
where $\bm{k}_S := \( \bm{k}_1 - \bm{k}_2 \) /2$. Taking a non-symmetric alternative for $\bm{k}_S$ would lead to $\Ord(k_L/k_S)$ terms for $\De B_{\ze_L XX}$, as is also the case for $B_{X_L\ze_1\ze_2}$ in the $\ze$-gauge \cite{Creminelli:2011rh,Pajer:2020wxk}. Note that the $\De B_{XYZ}$ contain no ``total energy poles'', i.e. terms proportional to $\( k_1 + k_2 + k_3 \)^{n<0}$. This is consistent with the argumentation of \cite{Pajer:2020wxk} that such bispectrum contributions are invariant under field redefinitions and thus gauge transformations in particular.

The above result is consistent with \eqref{eq:Bresult}, but the present derivation does not take into account the evaluation at horizon crossing time and thus does not allow us to determine whether the resulting $\be_*(k)$ is evaluated on $k_L$ or $k_S$. Indeed, as is well-known already from the classic computation \cite{Maldacena:2002vr}, the squeezed limit requires an alternative treatment that exploits the fact that long modes locally amount to `large' diffeomorphisms and is thus independent of the underlying action. We thus now perform this derivation, but with the real-space approach \cite{Creminelli:2004yq,Cheung:2007sv}. 

{\it Direct squeezed limit derivation} -- Having assumed a single-field model in unitary temporal gauge, all the degrees of freedom are in the 3-metric \eqref{eq:gauge}. For generic models the correlations are negligible when all modes are inside the horizon, so that the spectra are essentially generated during horizon-crossing and then freeze when all modes are out. In the squeezed bispectrum configuration the long mode crosses the horizon earlier, so it is already frozen and classical when the short modes do. Thus, the only surviving quantum correlation is between the two short modes, but in a background geometry that is deformed by the long mode. Concretely, if $k_3 \ll k_1 \approx k_2$, then \cite{Maldacena:2002vr,Creminelli:2004yq,Cheung:2007sv}
\beq \label{eq:SLCR}
B_{XYZ}(\bm{k}_1, \bm{k}_2, \bm{k}_3) \approx \bra \bra X_*(\bm{k}_1) \, Y_*(\bm{k}_2) \ket_{h_L} Z_*(\bm{k}_3) \ket'_{\rm cl.} \, ,
\eeq
where $\bra \dots \ket_{h_L}$ denotes the VEV on the background geometry deformed by the classical long-mode perturbation $h_{ij}(\bm{k}_L)$, while $\bra \dots \ket'_{\rm cl.}$ denotes a classical ensemble average without the Dirac delta. We start in real space, i.e. with the deformed 2-point correlation function $\bra X(\bm{x}_1) \, Y(\bm{x}_2) \ket_{h_L}$, so the squeezed configuration translates in $| \bm{x}_1 - \bm{x}_2 |$ being much smaller than the typical wave-length $\sim 1/k_L$ of the background geometry fluctuation. The effect of the long mode can then be treated as follows. First we have that
\bea
g_{ij} \ed x^i \ed x^j & = & a^2 \[ e^{2 \bm{h}_S + 2 \bm{h}_L} \]_{ij} \ed x^i \ed x^j \label{eq:BCH} \\
 & = & a^2 \[ e^{\bm{h}_L} e^{2 \bm{h}_S} e^{\bm{h}_L} + \Ord(\bm{h}^3) \]_{ij} \ed x^i \ed x^j \nn \\
 & \approx & a^2 \[ e^{2 \bm{h}_S} \]_{ij} \[ e^{\bm{h}_L} \]_{ik} \ed x^k \[ e^{\bm{h}_L} \]_{jl} \ed x^j \, , \nn 
\eea
and third-order terms do not influence the leading-order bispectrum. Next, we note that we can ignore the time-derivatives of $h_{ij,L}$ thanks to super-horizon conservation. Then, there exist spatial profiles $h_{ij,L}(\bm{x})$ which can be reabsorbed in the differentials \cite{Creminelli:2004yq,Cheung:2007sv,Creminelli:2012ed,Hinterbichler:2012nm}
\beq \label{eq:LD0} 
g_{ij}(\bm{x})\, \ed x^i \ed x^j = a^2 \[ e^{2 \bm{h}_S(\bm{x})} \]_{ij}\, \ed x'^i \ed x'^j \, ,
\eeq
where
\beq \label{eq:xtoxp}
x'^i = \[ e^{\bm{h}_L(\bm{x})} \]_{ij} x^j + \Ord(\pa h_L) \, .
\eeq
Moreover, if $h_{ij,S}$ transforms as a set of scalars under the diffeomorphism \eqref{eq:xtoxp}, then \eqref{eq:LD0} becomes
\beq \label{eq:LD} 
g_{ij}(\bm{x})\, \ed x^i \ed x^j = a^2 \[ e^{2 \bm{h}'_S(\bm{x}')} \]_{ij}\, \ed x'^i \ed x'^j \, ,
\eeq
meaning that those $h_{ij,L}(\bm{x})$ profiles can be eliminated with a diffeomorphism. These include the constant profile and, depending on the form of $h_{ij,S}$, also the linear ones $\sim \Ord(\bm{x})$ \cite{Creminelli:2012ed,Hinterbichler:2012nm}. But the latter would correspond to the $\Ord(\pa h_L)$ factors in \eqref{eq:xtoxp}, since they are absent for constant $h_{ij,L}$. Consequently, they lead to $\Ord(k_L)$ terms in Fourier space, so we can ignore them in the squeezed limit $k_L \to 0$ and we will come back to them later. Thus, the effect of the long mode in the deformed short-mode correlation function can be obtained by performing the large diffeomorphism $\bm{x} \to e^{\bm{h}_L(\bm{x})} \bm{x}$ on the undeformed one, up to $\Ord(k_L)$ corrections. For \eqref{eq:SLCR} we then only need the linear-order effect in $h_{ij,L}$
\bea
\bra X(\bm{x}_1) \, Y(\bm{x}_2) \ket_{h_L} & = & \bra X(e^{\bm{h}_L(\bm{x}_1)} \bm{x}_1) \, Y(e^{\bm{h}_L(\bm{x}_2)} \bm{x}_2) \ket \label{eq:LDcorr} \\
 & \approx & \bra X Y \ket(r) + h_{ij,L}(\bm{x})\, \hat{r}^i \hat{r}^j \, \frac{\ed \bra X Y \ket(r)}{\ed \log r} \,  , \nn
\eea
where we have defined the variables
\beq \label{eq:xrdefs}
\bm{x} := \frac{\bm{x}_1 + \bm{x}_2}{2} \, , \hspace{1cm} \bm{r} := \bm{x}_1 - \bm{x}_2 \, ,
\eeq
and used the fact that the free correlation function depends only on $r$. One could have also obtained \eqref{eq:LDcorr} by noting that the effect of the long mode is to simply alter that short distance $r^2 \to r^2 + 2 h_{ij,L}(\bm{x})\, r^i r^j$. The problem with this argument is that it is independent of the non-linear part of the 3-metric/fluctuation relation $g_{ij}[\bm{h}]$, while the bispectrum is sensitive to it. The precise derivation \eqref{eq:LD} is therefore necessary, because only the exponential parametrization \eqref{eq:pertdef} is able to factorize \cite{Maldacena:2002vr,Creminelli:2004yq}. We next go to Fourier space 
\bea
& & \bra X(\bm{k}_1) \, Y(\bm{k}_2) \ket_{h_L} \label{eq:FourCR} \\
& := & \int \ed^3 x_1\, \ed^3 x_2 \, e^{-i \bm{k}_1 \cdot \bm{x}_1 - i \bm{k}_2 \cdot \bm{x}_2} \bra X(\bm{x}_1) \, Y(\bm{x}_2) \ket_{h_L(\bm{x})}  \, , \nn
\eea
use $\pa_{\log r} e^{i \bm{k} \cdot \bm{r}} = \pa_{\log k} e^{i \bm{k} \cdot \bm{r}}$ and integrate by parts to find
\bea 
& & \bra X_*(\bm{k}_1) \, Y_*(\bm{k}_2) \ket_h \approx (2\pi)^3 \de^{(3)}(\bm{k}_L) \, P_{XY*}(k_S) \label{eq:FourCRint} \\
& & -\, h_{ij*}(\bm{k}_L) \int \ed^3 r \, \frac{\ed^3 k}{(2\pi)^3 k^2} \frac{\ed}{\ed k} \[ k^3 P_{XY*}(k) \] \hat{r}^i \hat{r}^j e^{i \( \bm{k} - \bm{k}_S \) \cdot \bm{r}} \, , \nn
\eea
where
\beq \label{eq:kLSofk123}
\bm{k}_L := \bm{k}_1 + \bm{k}_2 \, , \hspace{1cm} \bm{k}_S := \frac{1}{2} \( \bm{k}_1 - \bm{k}_2 \) \, , 
\eeq
are the long and (symmetric) short wave-vectors, if one takes into account the total momentum conservation. In the pure-scalar $\ze$-gauge case one has $h_{ij,L} \propto \de_{ij}$, so the integral over $r^i$ yields a Dirac delta, thus concluding the integration \cite{Creminelli:2004yq,Cheung:2007sv}. Here the structure of $h_{ij}$ is more general \eqref{eq:gauge}, so we do not have this simplification. In particular, we cannot switch the integration order in \eqref{eq:FourCRint} to perform the full $r^i$ integral, because this yields an infrared divergence. Instead, we can first integrate over the angles of $k^i$ and then over the angles of $r^i$ to find
\bea 
 & & \bra X_*(\bm{k}_1) \, Y_*(\bm{k}_2) \ket_h \approx (2\pi)^3 \de^{(3)}(\bm{k}_L) \, P_{XY*}(k_S) \nn \\
 & & -\, \frac{2}{\pi} \int_0^{\infty} r^2 \ed r \[ \( {\rm sinc} \,(k_S r) + \be_*(k_L) \, f(k_S r) \) \ti{\ze}_*(\bm{k}_L) \right. \nn \\
  & & \hspace{0.4cm} \left. + \, \( {\rm sinc}\,(k_S r) - 3 f(k_S r) \) \( \be_*(k_L)\, \ti{\ze}_*(\bm{k}_L) \, (\hat{\bm{k}}_L \cdot \hat{\bm{k}}_S)^2 \right. \right. \nn \\
  & & \hspace{4.4cm} \left. \left. +\, \ti{\ga}_{a*}(\bm{k}_L) \, e^a_{ij}(\hat{\bm{k}}_L)\, \hat{k}_S^i \hat{k}_S^j \) \] \nn \\
 & & \hspace{1cm} \times \int_0^{\infty} \ed k\, \frac{\ed}{\ed k} \[ k^3 P_{XY*}(k) \]{\rm sinc}\( k r \) \, , 
\eea
where $f(x) := \( {\rm sinc}\, x - \cos x \)/x^2$. We can now perform the integrals over $r$ using an exponential regulator
\bea
\lim_{\ep \to 0} \int_0^{\infty} \ed r\, r^2 \, {\rm sinc}\,(k r)\, {\rm sinc}\,(k_S r)\, e^{-\ep r} & = & \frac{\pi}{2k_S^2}\, \de\( k - k_S \) \, , \nn \\
\lim_{\ep \to 0} \int_0^{\infty} \ed r\, r^2 \, {\rm sinc}\,(k r) \, f(k_S r) \, e^{-\ep r} & = & \frac{\pi}{2 k_S^3} \, \te \( k_S - k \) \, ,  \nn 
\eea
so that, integrating over $k$, 
\bea 
& & \bra X_*(\bm{k}_1) \, Y_*(\bm{k}_2) \ket_h  \nn \\
& \approx & (2\pi)^3 \de^{(3)}(\bm{k}_L) \, P_{XY*}(k_S) \\
& & - \[ \frac{\ed \log \[ k_S^3 P_{XY*}(k_S) \]}{\ed \log k_S} + \be_*(k_L) \] P_{XY*}(k_S)\, \ti{\ze}_*(\bm{k}_L) \nn \\
& & - \, \frac{\ed \log P_{XY*}(k_S)}{\ed \log k_S} \, P_{XY*}(k_S) \nn \\
& & \times \[ \be_*(k_L)\, \ti{\ze}_*(\bm{k}_L) \, (\hat{\bm{k}}_L \cdot \hat{\bm{k}}_S)^2 + \ti{\ga}_{a*}(\bm{k}_L) \, e^a_{ij}(\hat{\bm{k}}_L)\, \hat{k}_S^i \hat{k}_S^j \] \, .  \nn
\eea
We can finally use \eqref{eq:SLCR} and \eqref{eq:Pkfree} to find the four non-trivial squeezed bispectra
\bea
\frac{B_{\ti{\ze}_L\ti{\ze}_1\ti{\ze}_2}}{P_{\ti{\ze}*}(k_L) \, P_{\ti{\ze}*}(k_S)} & \approx & - \[ n_{\ti{\ze}}(k_S) + \be_*(k_L) \]   \label{eq:BsssSLCR} \\
 & & + \, \be_*(k_L) \[ 3 - n_{\ti{\ze}}(k_S) \] (\hat{\bm{k}}_S \cdot \hat{\bm{k}}_L)^2 \, , \nn \\
\frac{B_{\ti{\ze}_L\ti{\ga}_{a,1}\ti{\ga}_{b,2}}}{P_{\ti{\ze}*}(k_L) \, P_{\ti{\ga}*}(k_S)} & \approx & - \[ n_{\ti{\ga}}(k_S) + \be_*(k_L) \] \de_{ab}  \label{eq:BsttSLCR} \\
 & & + \, \be_*(k_L) \[ 3 - n_{\ti{\ga}}(k_S) \] (\hat{\bm{k}}_S \cdot \hat{\bm{k}}_L)^2 \de_{ab} \, , \nn 
\eea
and
\bea
\frac{B_{\ti{\ga}_{a,L}\ti{\ze}_1\ti{\ze}_2}}{P_{\ti{\ga}*}(k_L) \, P_{\ti{\ze}*}(k_S)} & \approx & \[ 3 - n_{\ti{\ze}}(k_S) \] e_{a,L}^{ij} \hat{k}_S^i \hat{k}_S^j \, , \label{eq:BsstSLCR} \\
\frac{B_{\ti{\ga}_{a,L}\ti{\ga}_{b,1}\ti{\ga}_{c,2}}}{P_{\ti{\ga}*}(k_L) \, P_{\ti{\ga}*}(k_S)} & \approx & \[ 3 - n_{\ti{\ga}}(k_S) \] e_{a,L}^{ij} \hat{k}_S^i \hat{k}_S^j \de_{bc} \, . \label{eq:BtttSLCR}
\eea 
We have thus obtained \eqref{eq:Bresult}, with the $\be$ factor evaluated at the long mode crossing time. On top of this, we now also have the long tensor mode cases \eqref{eq:BsstSLCR} and \eqref{eq:BtttSLCR}, which are independent of $\be$. In particular, we check that all these expressions reproduce the classic result for slow-roll models \cite{Maldacena:2002vr} when $\be = 0$. Finally, we now see that the shape-dependence $\sim (\hat{\bm{k}}_S \cdot \hat{\bm{k}}_L)^2$ is nothing but the scalar analogue of the shape-dependence $\sim e_{a,L}^{ij} \hat{k}_S^i \hat{k}_S^j$ in the case of tensor long mode. Since they correspond to longitudinal and transverse anisotropies, respectively, they have the opposite behavior in shape space, i.e. the former vanishes in the isosceles configuration $\bm{k}_L \cdot \bm{k}_S = 0$, while the latter vanishes in the flatten one $\bm{k}_L \propto \bm{k}_S$. 

{\it Deviation from the squeezed limit} -- We are now interested in the finite-$k_L$ corrections to equations \eqref{eq:BsssSLCR} to \eqref{eq:BtttSLCR}, starting with what is known for the $\ze$-gauge ($\be = 0$) \cite{Creminelli:2004yq,Creminelli:2011rh,Maldacena:2011nz,Creminelli:2012ed,Hinterbichler:2012nm,Hinterbichler:2013dpa,Goldberger:2013rsa,Kundu:2014gxa,Kundu:2015xta,Shukla:2016bnu,Hui:2018cag,Pajer:2020wxk}. If the short modes are scalars, then the short-mode 3-metric is conformally flat $g_{ij,S} = a^2 e^{2\ze_S} \de_{ij}$ and this implies that one can also eliminate the linear profiles $h_{ij,L} \sim \Ord(\bm{x})$ through a large diffeomorphism \eqref{eq:LD} at $\Ord(h_L)$, which is a special conformal transformation in the long scalar mode case \cite{Creminelli:2012ed,Hinterbichler:2012nm}. The elimination of the constant mode meant that we could implement its effect on the short-mode power spectrum as a large spatial diffeomorphism and this led to the consistency relation in the exact $k_L = 0$ case. Similarly, the elimination of the $\Ord(x\pa h_L)$ terms leads to extra large spatial diffeomorphisms and thus to a consistency relation for the $\Ord(k_L/k_S)$ bispectrum, the so-called `conformal consistency relation' \cite{Creminelli:2012ed,Hinterbichler:2013dpa}. However, contrary to \eqref{eq:LDcorr}, the corresponding diffeomorphisms do not alter the short-mode 2-point function $\bra X_S Y_S \ket(r)$ at $\Ord(h_L)$, so there is no $\Ord(k_L/k_S)$ correction for short scalar modes in the $\ze$-gauge.\footnote{Again, this only holds if $\bm{k}_S$ is the symmetric average \eqref{eq:kLSofk123} \cite{Creminelli:2011rh,Pajer:2020wxk}.} 
Finally, if the short modes are tensors, then the short-mode 3-metric is not conformally flat and this obstructs the elimination of non-trivial long-mode profiles through large diffeomorphisms \cite{Creminelli:2012ed,Hinterbichler:2012nm}. The corrections therefore start at $\Ord(k_L/k_S)$.  

In the $\be \neq 0$ case we are interested in here, equations \eqref{eq:DeBsssSq} and \eqref{eq:DeBtssSq} show that there is still no $\Ord(k_L/k_S)$ correction for the case of scalar short modes. To understand why this is the case, note that $g_{ij,S}$ is related to $a^2 e^{2 \ze_S} \de_{ij}$ through a spatial gauge transformation, so we can perform that transformation to reach the conformally flat 3-metric, then the extra large diffeomorphisms and finally come back with the inverse gauge transformation. The number of extra large spatial diffeomorphisms is therefore the same for all $\be$, hence our result (the same argument applies to the CFC case \cite{Pajer:2013ana}). Finally, in the case of the particular gauge \eqref{eq:be}, note that the passage to \eqref{eq:Brun} introduces a $\sim \al_X(k_S)\, k_L/k_S$ correction.

\begin{acknowledgments}
We are very grateful to Enrico Pajer for useful remarks and suggestions and also thank Toni Riotto and Matias Zaldarriaga. This work is supported by a Consolidator Grant of the European Research Council (ERC-2015-CoG grant 680886).
\end{acknowledgments}

\bibliography{mybib}

\end{document}

%% file: mydefs.tex
\newcommand{\beq}{\begin{equation}}
\newcommand{\eeq}{\end{equation}}
\newcommand{\bea}{\begin{eqnarray}}
\newcommand{\eea}{\end{eqnarray}}
\newcommand{\ben}{\begin{enumerate}}
\newcommand{\een}{\end{enumerate}}

\newcommand{\pa}{\partial}

\newcommand{\ed}{{\rm d}}

\newcommand{\Ord}{{\cal O}}

\newcommand{\ti}{\tilde}

\renewcommand\({\left(}
\renewcommand\){\right)}
\renewcommand\[{\left[}
\renewcommand\]{\right]}
\newcommand{\bra}{\langle}
\newcommand{\ket}{\rangle}

\newcommand{\os}{\overset}

\newcommand{\nn}{\nonumber}

\newcommand{\al}{\alpha}
\newcommand{\be}{\beta}
\newcommand{\ga}{\gamma}

\newcommand{\de}{\delta}
\newcommand{\De}{\Delta}
\newcommand{\ep}{\epsilon}
\newcommand{\vep}{\varepsilon}
\newcommand{\ze}{\zeta}
\newcommand{\et}{\eta}
\newcommand{\te}{\theta}

\newcommand{\cH}{{\cal H}}

%% file: NSB_let_revised.bbl
\begin{thebibliography}{68}%
\makeatletter
\providecommand \@ifxundefined [1]{%
 \@ifx{#1\undefined}
}%
\providecommand \@ifnum [1]{%
 \ifnum #1\expandafter \@firstoftwo
 \else \expandafter \@secondoftwo
 \fi
}%
\providecommand \@ifx [1]{%
 \ifx #1\expandafter \@firstoftwo
 \else \expandafter \@secondoftwo
 \fi
}%
\providecommand \natexlab [1]{#1}%
\providecommand \enquote  [1]{``#1''}%
\providecommand \bibnamefont  [1]{#1}%
\providecommand \bibfnamefont [1]{#1}%
\providecommand \citenamefont [1]{#1}%
\providecommand \href@noop [0]{\@secondoftwo}%
\providecommand \href [0]{\begingroup \@sanitize@url \@href}%
\providecommand \@href[1]{\@@startlink{#1}\@@href}%
\providecommand \@@href[1]{\endgroup#1\@@endlink}%
\providecommand \@sanitize@url [0]{\catcode `\\12\catcode `\$12\catcode
  `\&12\catcode `\#12\catcode `\^12\catcode `\_12\catcode `\%12\relax}%
\providecommand \@@startlink[1]{}%
\providecommand \@@endlink[0]{}%
\providecommand \url  [0]{\begingroup\@sanitize@url \@url }%
\providecommand \@url [1]{\endgroup\@href {#1}{\urlprefix }}%
\providecommand \urlprefix  [0]{URL }%
\providecommand \Eprint [0]{\href }%
\providecommand \doibase [0]{https://doi.org/}%
\providecommand \selectlanguage [0]{\@gobble}%
\providecommand \bibinfo  [0]{\@secondoftwo}%
\providecommand \bibfield  [0]{\@secondoftwo}%
\providecommand \translation [1]{[#1]}%
\providecommand \BibitemOpen [0]{}%
\providecommand \bibitemStop [0]{}%
\providecommand \bibitemNoStop [0]{.\EOS\space}%
\providecommand \EOS [0]{\spacefactor3000\relax}%
\providecommand \BibitemShut  [1]{\csname bibitem#1\endcsname}%
\let\auto@bib@innerbib\@empty
\bibitem [{\citenamefont {Dewdney}\ \emph {et~al.}(2009)\citenamefont
  {Dewdney}, \citenamefont {Hall}, \citenamefont {Schilizzi},\ and\
  \citenamefont {Lazio}}]{SKA}%
  \BibitemOpen
  \bibfield  {author} {\bibinfo {author} {\bibfnamefont {P.~E.}\ \bibnamefont
  {Dewdney}}, \bibinfo {author} {\bibfnamefont {P.~J.}\ \bibnamefont {Hall}},
  \bibinfo {author} {\bibfnamefont {R.~T.}\ \bibnamefont {Schilizzi}},\ and\
  \bibinfo {author} {\bibfnamefont {T.~J. L.~W.}\ \bibnamefont {Lazio}},\
  }\bibfield  {title} {\bibinfo {title} {{The Square Kilometre Array}},\ }\href
  {https://doi.org/10.1109/JPROC.2009.2021005} {\bibfield  {journal} {\bibinfo
  {journal} {Proceedings of the IEEE}\ }\textbf {\bibinfo {volume} {97}},\
  \bibinfo {pages} {1482} (\bibinfo {year} {2009})}\BibitemShut {NoStop}%
\bibitem [{\citenamefont {Laureijs}\ \emph {et~al.}(2011)\citenamefont
  {Laureijs} \emph {et~al.}}]{EUCLID}%
  \BibitemOpen
  \bibfield  {author} {\bibinfo {author} {\bibfnamefont {R.}~\bibnamefont
  {Laureijs}} \emph {et~al.},\ }\bibfield  {title} {\bibinfo {title} {{Euclid
  Definition Study Report}},\ }\href@noop {} {\  (\bibinfo {year} {2011})},\
  \Eprint {https://arxiv.org/abs/1110.3193} {arXiv:1110.3193} \BibitemShut
  {NoStop}%
\bibitem [{\citenamefont {Green}\ \emph {et~al.}(2012)\citenamefont {Green}
  \emph {et~al.}}]{WFIRST}%
  \BibitemOpen
  \bibfield  {author} {\bibinfo {author} {\bibfnamefont {J.}~\bibnamefont
  {Green}} \emph {et~al.},\ }\bibfield  {title} {\bibinfo {title} {{Wide-Field
  InfraRed Survey Telescope (WFIRST) Final Report}},\ }\href@noop {} {\
  (\bibinfo {year} {2012})},\ \Eprint {https://arxiv.org/abs/1208.4012}
  {arXiv:1208.4012} \BibitemShut {NoStop}%
\bibitem [{\citenamefont {Levi}\ \emph {et~al.}(2013)\citenamefont {Levi} \emph
  {et~al.}}]{DESI}%
  \BibitemOpen
  \bibfield  {author} {\bibinfo {author} {\bibfnamefont {M.}~\bibnamefont
  {Levi}} \emph {et~al.},\ }\bibfield  {title} {\bibinfo {title} {{The DESI
  Experiment, a whitepaper for Snowmass 2013}},\ }\href@noop {} {\  (\bibinfo
  {year} {2013})},\ \Eprint {https://arxiv.org/abs/1308.0847} {arXiv:1308.0847}
  \BibitemShut {NoStop}%
\bibitem [{\citenamefont {Ivezic}\ \emph {et~al.}(2019)\citenamefont {Ivezic}
  \emph {et~al.}}]{LSST}%
  \BibitemOpen
  \bibfield  {author} {\bibinfo {author} {\bibfnamefont {Z.}~\bibnamefont
  {Ivezic}} \emph {et~al.},\ }\bibfield  {title} {\bibinfo {title} {{LSST: From
  Science Drivers to Reference Design and Anticipated Data Products}},\ }\href
  {https://doi.org/10.3847/1538-4357/ab042c} {\bibfield  {journal} {\bibinfo
  {journal} {The Astrophysical Journal}\ }\textbf {\bibinfo {volume} {873}},\
  \bibinfo {pages} {111} (\bibinfo {year} {2019})}\BibitemShut {NoStop}%
\bibitem [{\citenamefont {Dalal}\ \emph {et~al.}(2008)\citenamefont {Dalal},
  \citenamefont {Dore}, \citenamefont {Huterer},\ and\ \citenamefont
  {Shirokov}}]{Dalal:2007cu}%
  \BibitemOpen
  \bibfield  {author} {\bibinfo {author} {\bibfnamefont {N.}~\bibnamefont
  {Dalal}}, \bibinfo {author} {\bibfnamefont {O.}~\bibnamefont {Dore}},
  \bibinfo {author} {\bibfnamefont {D.}~\bibnamefont {Huterer}},\ and\ \bibinfo
  {author} {\bibfnamefont {A.}~\bibnamefont {Shirokov}},\ }\bibfield  {title}
  {\bibinfo {title} {{The imprints of primordial non-gaussianities on
  large-scale structure: scale dependent bias and abundance of virialized
  objects}},\ }\href {https://doi.org/10.1103/PhysRevD.77.123514} {\bibfield
  {journal} {\bibinfo  {journal} {Phys. Rev. D}\ }\textbf {\bibinfo {volume}
  {77}},\ \bibinfo {pages} {123514} (\bibinfo {year} {2008})},\ \Eprint
  {https://arxiv.org/abs/0710.4560} {arXiv:0710.4560 [astro-ph]} \BibitemShut
  {NoStop}%
\bibitem [{\citenamefont {Matarrese}\ and\ \citenamefont
  {Verde}(2008)}]{Matarrese:2008nc}%
  \BibitemOpen
  \bibfield  {author} {\bibinfo {author} {\bibfnamefont {S.}~\bibnamefont
  {Matarrese}}\ and\ \bibinfo {author} {\bibfnamefont {L.}~\bibnamefont
  {Verde}},\ }\bibfield  {title} {\bibinfo {title} {{The effect of primordial
  non-Gaussianity on halo bias}},\ }\href {https://doi.org/10.1086/587840}
  {\bibfield  {journal} {\bibinfo  {journal} {Astrophys. J. Lett.}\ }\textbf
  {\bibinfo {volume} {677}},\ \bibinfo {pages} {L77} (\bibinfo {year}
  {2008})},\ \Eprint {https://arxiv.org/abs/0801.4826} {arXiv:0801.4826
  [astro-ph]} \BibitemShut {NoStop}%
\bibitem [{\citenamefont {Maldacena}(2003)}]{Maldacena:2002vr}%
  \BibitemOpen
  \bibfield  {author} {\bibinfo {author} {\bibfnamefont {J.~M.}\ \bibnamefont
  {Maldacena}},\ }\bibfield  {title} {\bibinfo {title} {{Non-Gaussian features
  of primordial fluctuations in single field inflationary models}},\ }\href
  {https://doi.org/10.1088/1126-6708/2003/05/013} {\bibfield  {journal}
  {\bibinfo  {journal} {JHEP}\ }\textbf {\bibinfo {volume} {05}},\ \bibinfo
  {pages} {013}},\ \Eprint {https://arxiv.org/abs/astro-ph/0210603}
  {arXiv:astro-ph/0210603} \BibitemShut {NoStop}%
\bibitem [{\citenamefont {Creminelli}\ and\ \citenamefont
  {Zaldarriaga}(2004)}]{Creminelli:2004yq}%
  \BibitemOpen
  \bibfield  {author} {\bibinfo {author} {\bibfnamefont {P.}~\bibnamefont
  {Creminelli}}\ and\ \bibinfo {author} {\bibfnamefont {M.}~\bibnamefont
  {Zaldarriaga}},\ }\bibfield  {title} {\bibinfo {title} {{Single field
  consistency relation for the 3-point function}},\ }\href
  {https://doi.org/10.1088/1475-7516/2004/10/006} {\bibfield  {journal}
  {\bibinfo  {journal} {JCAP}\ }\textbf {\bibinfo {volume} {10}},\ \bibinfo
  {pages} {006}},\ \Eprint {https://arxiv.org/abs/astro-ph/0407059}
  {arXiv:astro-ph/0407059} \BibitemShut {NoStop}%
\bibitem [{\citenamefont {Cheung}\ \emph {et~al.}(2008)\citenamefont {Cheung},
  \citenamefont {Fitzpatrick}, \citenamefont {Kaplan},\ and\ \citenamefont
  {Senatore}}]{Cheung:2007sv}%
  \BibitemOpen
  \bibfield  {author} {\bibinfo {author} {\bibfnamefont {C.}~\bibnamefont
  {Cheung}}, \bibinfo {author} {\bibfnamefont {A.~L.}\ \bibnamefont
  {Fitzpatrick}}, \bibinfo {author} {\bibfnamefont {J.}~\bibnamefont
  {Kaplan}},\ and\ \bibinfo {author} {\bibfnamefont {L.}~\bibnamefont
  {Senatore}},\ }\bibfield  {title} {\bibinfo {title} {{On the consistency
  relation of the 3-point function in single field inflation}},\ }\href
  {https://doi.org/10.1088/1475-7516/2008/02/021} {\bibfield  {journal}
  {\bibinfo  {journal} {JCAP}\ }\textbf {\bibinfo {volume} {02}},\ \bibinfo
  {pages} {021}},\ \Eprint {https://arxiv.org/abs/0709.0295} {arXiv:0709.0295
  [hep-th]} \BibitemShut {NoStop}%
\bibitem [{\citenamefont {Ganc}\ and\ \citenamefont
  {Komatsu}(2010)}]{Ganc:2010ff}%
  \BibitemOpen
  \bibfield  {author} {\bibinfo {author} {\bibfnamefont {J.}~\bibnamefont
  {Ganc}}\ and\ \bibinfo {author} {\bibfnamefont {E.}~\bibnamefont {Komatsu}},\
  }\bibfield  {title} {\bibinfo {title} {{A new method for calculating the
  primordial bispectrum in the squeezed limit}},\ }\href
  {https://doi.org/10.1088/1475-7516/2010/12/009} {\bibfield  {journal}
  {\bibinfo  {journal} {JCAP}\ }\textbf {\bibinfo {volume} {12}},\ \bibinfo
  {pages} {009}},\ \Eprint {https://arxiv.org/abs/1006.5457} {arXiv:1006.5457
  [astro-ph.CO]} \BibitemShut {NoStop}%
\bibitem [{\citenamefont {Creminelli}\ \emph {et~al.}(2011)\citenamefont
  {Creminelli}, \citenamefont {D'Amico}, \citenamefont {Musso},\ and\
  \citenamefont {Norena}}]{Creminelli:2011rh}%
  \BibitemOpen
  \bibfield  {author} {\bibinfo {author} {\bibfnamefont {P.}~\bibnamefont
  {Creminelli}}, \bibinfo {author} {\bibfnamefont {G.}~\bibnamefont {D'Amico}},
  \bibinfo {author} {\bibfnamefont {M.}~\bibnamefont {Musso}},\ and\ \bibinfo
  {author} {\bibfnamefont {J.}~\bibnamefont {Norena}},\ }\bibfield  {title}
  {\bibinfo {title} {{The (not so) squeezed limit of the primordial 3-point
  function}},\ }\href {https://doi.org/10.1088/1475-7516/2011/11/038}
  {\bibfield  {journal} {\bibinfo  {journal} {JCAP}\ }\textbf {\bibinfo
  {volume} {11}},\ \bibinfo {pages} {038}},\ \Eprint
  {https://arxiv.org/abs/1106.1462} {arXiv:1106.1462 [astro-ph.CO]}
  \BibitemShut {NoStop}%
\bibitem [{\citenamefont {Pimentel}(2014)}]{Pimentel:2013gza}%
  \BibitemOpen
  \bibfield  {author} {\bibinfo {author} {\bibfnamefont {G.~L.}\ \bibnamefont
  {Pimentel}},\ }\bibfield  {title} {\bibinfo {title} {{Inflationary
  Consistency Conditions from a Wavefunctional Perspective}},\ }\href
  {https://doi.org/10.1007/JHEP02(2014)124} {\bibfield  {journal} {\bibinfo
  {journal} {JHEP}\ }\textbf {\bibinfo {volume} {02}},\ \bibinfo {pages}
  {124}},\ \Eprint {https://arxiv.org/abs/1309.1793} {arXiv:1309.1793 [hep-th]}
  \BibitemShut {NoStop}%
\bibitem [{\citenamefont {Kinney}(2005)}]{Kinney:2005vj}%
  \BibitemOpen
  \bibfield  {author} {\bibinfo {author} {\bibfnamefont {W.~H.}\ \bibnamefont
  {Kinney}},\ }\bibfield  {title} {\bibinfo {title} {{Horizon crossing and
  inflation with large eta}},\ }\href
  {https://doi.org/10.1103/PhysRevD.72.023515} {\bibfield  {journal} {\bibinfo
  {journal} {Phys. Rev. D}\ }\textbf {\bibinfo {volume} {72}},\ \bibinfo
  {pages} {023515} (\bibinfo {year} {2005})},\ \Eprint
  {https://arxiv.org/abs/gr-qc/0503017} {arXiv:gr-qc/0503017} \BibitemShut
  {NoStop}%
\bibitem [{\citenamefont {Chen}(2010)}]{Chen:2010xka}%
  \BibitemOpen
  \bibfield  {author} {\bibinfo {author} {\bibfnamefont {X.}~\bibnamefont
  {Chen}},\ }\bibfield  {title} {\bibinfo {title} {{Primordial
  Non-Gaussianities from Inflation Models}},\ }\href
  {https://doi.org/10.1155/2010/638979} {\bibfield  {journal} {\bibinfo
  {journal} {Adv. Astron.}\ }\textbf {\bibinfo {volume} {2010}},\ \bibinfo
  {pages} {638979} (\bibinfo {year} {2010})},\ \Eprint
  {https://arxiv.org/abs/1002.1416} {arXiv:1002.1416 [astro-ph.CO]}
  \BibitemShut {NoStop}%
\bibitem [{\citenamefont {Namjoo}\ \emph {et~al.}(2013)\citenamefont {Namjoo},
  \citenamefont {Firouzjahi},\ and\ \citenamefont {Sasaki}}]{Namjoo:2012aa}%
  \BibitemOpen
  \bibfield  {author} {\bibinfo {author} {\bibfnamefont {M.~H.}\ \bibnamefont
  {Namjoo}}, \bibinfo {author} {\bibfnamefont {H.}~\bibnamefont {Firouzjahi}},\
  and\ \bibinfo {author} {\bibfnamefont {M.}~\bibnamefont {Sasaki}},\
  }\bibfield  {title} {\bibinfo {title} {{Violation of non-Gaussianity
  consistency relation in a single field inflationary model}},\ }\href
  {https://doi.org/10.1209/0295-5075/101/39001} {\bibfield  {journal} {\bibinfo
   {journal} {EPL}\ }\textbf {\bibinfo {volume} {101}},\ \bibinfo {pages}
  {39001} (\bibinfo {year} {2013})},\ \Eprint {https://arxiv.org/abs/1210.3692}
  {arXiv:1210.3692 [astro-ph.CO]} \BibitemShut {NoStop}%
\bibitem [{\citenamefont {Martin}\ \emph {et~al.}(2013)\citenamefont {Martin},
  \citenamefont {Motohashi},\ and\ \citenamefont {Suyama}}]{Martin:2012pe}%
  \BibitemOpen
  \bibfield  {author} {\bibinfo {author} {\bibfnamefont {J.}~\bibnamefont
  {Martin}}, \bibinfo {author} {\bibfnamefont {H.}~\bibnamefont {Motohashi}},\
  and\ \bibinfo {author} {\bibfnamefont {T.}~\bibnamefont {Suyama}},\
  }\bibfield  {title} {\bibinfo {title} {{Ultra Slow-Roll Inflation and the
  non-Gaussianity Consistency Relation}},\ }\href
  {https://doi.org/10.1103/PhysRevD.87.023514} {\bibfield  {journal} {\bibinfo
  {journal} {Phys. Rev. D}\ }\textbf {\bibinfo {volume} {87}},\ \bibinfo
  {pages} {023514} (\bibinfo {year} {2013})},\ \Eprint
  {https://arxiv.org/abs/1211.0083} {arXiv:1211.0083 [astro-ph.CO]}
  \BibitemShut {NoStop}%
\bibitem [{\citenamefont {Chen}\ \emph {et~al.}(2013)\citenamefont {Chen},
  \citenamefont {Firouzjahi}, \citenamefont {Namjoo},\ and\ \citenamefont
  {Sasaki}}]{Chen:2013aj}%
  \BibitemOpen
  \bibfield  {author} {\bibinfo {author} {\bibfnamefont {X.}~\bibnamefont
  {Chen}}, \bibinfo {author} {\bibfnamefont {H.}~\bibnamefont {Firouzjahi}},
  \bibinfo {author} {\bibfnamefont {M.~H.}\ \bibnamefont {Namjoo}},\ and\
  \bibinfo {author} {\bibfnamefont {M.}~\bibnamefont {Sasaki}},\ }\bibfield
  {title} {\bibinfo {title} {{A Single Field Inflation Model with Large Local
  Non-Gaussianity}},\ }\href {https://doi.org/10.1209/0295-5075/102/59001}
  {\bibfield  {journal} {\bibinfo  {journal} {EPL}\ }\textbf {\bibinfo {volume}
  {102}},\ \bibinfo {pages} {59001} (\bibinfo {year} {2013})},\ \Eprint
  {https://arxiv.org/abs/1301.5699} {arXiv:1301.5699 [hep-th]} \BibitemShut
  {NoStop}%
\bibitem [{\citenamefont {Mooij}\ and\ \citenamefont
  {Palma}(2015)}]{Mooij:2015yka}%
  \BibitemOpen
  \bibfield  {author} {\bibinfo {author} {\bibfnamefont {S.}~\bibnamefont
  {Mooij}}\ and\ \bibinfo {author} {\bibfnamefont {G.~A.}\ \bibnamefont
  {Palma}},\ }\bibfield  {title} {\bibinfo {title} {{Consistently violating the
  non-Gaussian consistency relation}},\ }\href
  {https://doi.org/10.1088/1475-7516/2015/11/025} {\bibfield  {journal}
  {\bibinfo  {journal} {JCAP}\ }\textbf {\bibinfo {volume} {11}},\ \bibinfo
  {pages} {025}},\ \Eprint {https://arxiv.org/abs/1502.03458} {arXiv:1502.03458
  [astro-ph.CO]} \BibitemShut {NoStop}%
\bibitem [{\citenamefont {Suyama}\ \emph {et~al.}(2021)\citenamefont {Suyama},
  \citenamefont {Tada},\ and\ \citenamefont {Yamaguchi}}]{Suyama:2021adn}%
  \BibitemOpen
  \bibfield  {author} {\bibinfo {author} {\bibfnamefont {T.}~\bibnamefont
  {Suyama}}, \bibinfo {author} {\bibfnamefont {Y.}~\bibnamefont {Tada}},\ and\
  \bibinfo {author} {\bibfnamefont {M.}~\bibnamefont {Yamaguchi}},\ }\bibfield
  {title} {\bibinfo {title} {{Revisiting non-Gaussianity in non-attractor
  inflation models in the light of the cosmological soft theorem}},\ }\href
  {https://doi.org/10.1093/ptep/ptab063} {\bibfield  {journal} {\bibinfo
  {journal} {PTEP}\ }\textbf {\bibinfo {volume} {2021}},\ \bibinfo {pages}
  {073E02} (\bibinfo {year} {2021})},\ \Eprint
  {https://arxiv.org/abs/2101.10682} {arXiv:2101.10682 [hep-th]} \BibitemShut
  {NoStop}%
\bibitem [{\citenamefont {Avis}\ \emph {et~al.}(2020)\citenamefont {Avis},
  \citenamefont {Jazayeri}, \citenamefont {Pajer},\ and\ \citenamefont
  {Supe\l{}}}]{Avis:2019eav}%
  \BibitemOpen
  \bibfield  {author} {\bibinfo {author} {\bibfnamefont {G.}~\bibnamefont
  {Avis}}, \bibinfo {author} {\bibfnamefont {S.}~\bibnamefont {Jazayeri}},
  \bibinfo {author} {\bibfnamefont {E.}~\bibnamefont {Pajer}},\ and\ \bibinfo
  {author} {\bibfnamefont {J.}~\bibnamefont {Supe\l{}}},\ }\bibfield  {title}
  {\bibinfo {title} {{Spatial Curvature at the Sound Horizon}},\ }\href
  {https://doi.org/10.1088/1475-7516/2020/02/034} {\bibfield  {journal}
  {\bibinfo  {journal} {JCAP}\ }\textbf {\bibinfo {volume} {02}},\ \bibinfo
  {pages} {034}},\ \Eprint {https://arxiv.org/abs/1911.04454} {arXiv:1911.04454
  [astro-ph.CO]} \BibitemShut {NoStop}%
\bibitem [{\citenamefont {Bravo}\ \emph
  {et~al.}(2018{\natexlab{a}})\citenamefont {Bravo}, \citenamefont {Mooij},
  \citenamefont {Palma},\ and\ \citenamefont {Pradenas}}]{Bravo:2017wyw}%
  \BibitemOpen
  \bibfield  {author} {\bibinfo {author} {\bibfnamefont {R.}~\bibnamefont
  {Bravo}}, \bibinfo {author} {\bibfnamefont {S.}~\bibnamefont {Mooij}},
  \bibinfo {author} {\bibfnamefont {G.~A.}\ \bibnamefont {Palma}},\ and\
  \bibinfo {author} {\bibfnamefont {B.}~\bibnamefont {Pradenas}},\ }\bibfield
  {title} {\bibinfo {title} {{A generalized non-Gaussian consistency relation
  for single field inflation}},\ }\href
  {https://doi.org/10.1088/1475-7516/2018/05/024} {\bibfield  {journal}
  {\bibinfo  {journal} {JCAP}\ }\textbf {\bibinfo {volume} {05}},\ \bibinfo
  {pages} {024}},\ \Eprint {https://arxiv.org/abs/1711.02680} {arXiv:1711.02680
  [astro-ph.CO]} \BibitemShut {NoStop}%
\bibitem [{\citenamefont {Finelli}\ \emph
  {et~al.}(2018{\natexlab{a}})\citenamefont {Finelli}, \citenamefont {Goon},
  \citenamefont {Pajer},\ and\ \citenamefont {Santoni}}]{Finelli:2017fml}%
  \BibitemOpen
  \bibfield  {author} {\bibinfo {author} {\bibfnamefont {B.}~\bibnamefont
  {Finelli}}, \bibinfo {author} {\bibfnamefont {G.}~\bibnamefont {Goon}},
  \bibinfo {author} {\bibfnamefont {E.}~\bibnamefont {Pajer}},\ and\ \bibinfo
  {author} {\bibfnamefont {L.}~\bibnamefont {Santoni}},\ }\bibfield  {title}
  {\bibinfo {title} {{Soft Theorems For Shift-Symmetric Cosmologies}},\ }\href
  {https://doi.org/10.1103/PhysRevD.97.063531} {\bibfield  {journal} {\bibinfo
  {journal} {Phys. Rev. D}\ }\textbf {\bibinfo {volume} {97}},\ \bibinfo
  {pages} {063531} (\bibinfo {year} {2018}{\natexlab{a}})},\ \Eprint
  {https://arxiv.org/abs/1711.03737} {arXiv:1711.03737 [hep-th]} \BibitemShut
  {NoStop}%
\bibitem [{\citenamefont {Finelli}\ \emph
  {et~al.}(2018{\natexlab{b}})\citenamefont {Finelli}, \citenamefont {Goon},
  \citenamefont {Pajer},\ and\ \citenamefont {Santoni}}]{Finelli:2018upr}%
  \BibitemOpen
  \bibfield  {author} {\bibinfo {author} {\bibfnamefont {B.}~\bibnamefont
  {Finelli}}, \bibinfo {author} {\bibfnamefont {G.}~\bibnamefont {Goon}},
  \bibinfo {author} {\bibfnamefont {E.}~\bibnamefont {Pajer}},\ and\ \bibinfo
  {author} {\bibfnamefont {L.}~\bibnamefont {Santoni}},\ }\bibfield  {title}
  {\bibinfo {title} {{The Effective Theory of Shift-Symmetric Cosmologies}},\
  }\href {https://doi.org/10.1088/1475-7516/2018/05/060} {\bibfield  {journal}
  {\bibinfo  {journal} {JCAP}\ }\textbf {\bibinfo {volume} {05}},\ \bibinfo
  {pages} {060}},\ \Eprint {https://arxiv.org/abs/1802.01580} {arXiv:1802.01580
  [hep-th]} \BibitemShut {NoStop}%
\bibitem [{\citenamefont {Bravo}\ and\ \citenamefont
  {Palma}(2020)}]{Bravo:2020hde}%
  \BibitemOpen
  \bibfield  {author} {\bibinfo {author} {\bibfnamefont {R.}~\bibnamefont
  {Bravo}}\ and\ \bibinfo {author} {\bibfnamefont {G.~A.}\ \bibnamefont
  {Palma}},\ }\bibfield  {title} {\bibinfo {title} {{Unifying attractor and
  non-attractor models of inflation under a single soft theorem}},\ }\href@noop
  {} {\  (\bibinfo {year} {2020})},\ \Eprint {https://arxiv.org/abs/2009.03369}
  {arXiv:2009.03369 [hep-th]} \BibitemShut {NoStop}%
\bibitem [{\citenamefont {Chen}\ \emph {et~al.}(2006)\citenamefont {Chen},
  \citenamefont {Huang},\ and\ \citenamefont {Shiu}}]{Chen:2006dfn}%
  \BibitemOpen
  \bibfield  {author} {\bibinfo {author} {\bibfnamefont {X.}~\bibnamefont
  {Chen}}, \bibinfo {author} {\bibfnamefont {M.-x.}\ \bibnamefont {Huang}},\
  and\ \bibinfo {author} {\bibfnamefont {G.}~\bibnamefont {Shiu}},\ }\bibfield
  {title} {\bibinfo {title} {{The Inflationary Trispectrum for Models with
  Large Non-Gaussianities}},\ }\href
  {https://doi.org/10.1103/PhysRevD.74.121301} {\bibfield  {journal} {\bibinfo
  {journal} {Phys. Rev. D}\ }\textbf {\bibinfo {volume} {74}},\ \bibinfo
  {pages} {121301} (\bibinfo {year} {2006})},\ \Eprint
  {https://arxiv.org/abs/hep-th/0610235} {arXiv:hep-th/0610235} \BibitemShut
  {NoStop}%
\bibitem [{\citenamefont {Creminelli}\ \emph {et~al.}(2012)\citenamefont
  {Creminelli}, \citenamefont {Nore\~na},\ and\ \citenamefont
  {Simonovi\'c}}]{Creminelli:2012ed}%
  \BibitemOpen
  \bibfield  {author} {\bibinfo {author} {\bibfnamefont {P.}~\bibnamefont
  {Creminelli}}, \bibinfo {author} {\bibfnamefont {J.}~\bibnamefont
  {Nore\~na}},\ and\ \bibinfo {author} {\bibfnamefont {M.}~\bibnamefont
  {Simonovi\'c}},\ }\bibfield  {title} {\bibinfo {title} {{Conformal
  consistency relations for single-field inflation}},\ }\href
  {https://doi.org/10.1088/1475-7516/2012/07/052} {\bibfield  {journal}
  {\bibinfo  {journal} {JCAP}\ }\textbf {\bibinfo {volume} {07}},\ \bibinfo
  {pages} {052}},\ \Eprint {https://arxiv.org/abs/1203.4595} {arXiv:1203.4595
  [hep-th]} \BibitemShut {NoStop}%
\bibitem [{\citenamefont {Senatore}\ and\ \citenamefont
  {Zaldarriaga}(2012)}]{Senatore:2012wy}%
  \BibitemOpen
  \bibfield  {author} {\bibinfo {author} {\bibfnamefont {L.}~\bibnamefont
  {Senatore}}\ and\ \bibinfo {author} {\bibfnamefont {M.}~\bibnamefont
  {Zaldarriaga}},\ }\bibfield  {title} {\bibinfo {title} {{A Note on the
  Consistency Condition of Primordial Fluctuations}},\ }\href
  {https://doi.org/10.1088/1475-7516/2012/08/001} {\bibfield  {journal}
  {\bibinfo  {journal} {JCAP}\ }\textbf {\bibinfo {volume} {08}},\ \bibinfo
  {pages} {001}},\ \Eprint {https://arxiv.org/abs/1203.6884} {arXiv:1203.6884
  [astro-ph.CO]} \BibitemShut {NoStop}%
\bibitem [{\citenamefont {Hinterbichler}\ \emph {et~al.}(2012)\citenamefont
  {Hinterbichler}, \citenamefont {Hui},\ and\ \citenamefont
  {Khoury}}]{Hinterbichler:2012nm}%
  \BibitemOpen
  \bibfield  {author} {\bibinfo {author} {\bibfnamefont {K.}~\bibnamefont
  {Hinterbichler}}, \bibinfo {author} {\bibfnamefont {L.}~\bibnamefont {Hui}},\
  and\ \bibinfo {author} {\bibfnamefont {J.}~\bibnamefont {Khoury}},\
  }\bibfield  {title} {\bibinfo {title} {{Conformal Symmetries of Adiabatic
  Modes in Cosmology}},\ }\href {https://doi.org/10.1088/1475-7516/2012/08/017}
  {\bibfield  {journal} {\bibinfo  {journal} {JCAP}\ }\textbf {\bibinfo
  {volume} {08}},\ \bibinfo {pages} {017}},\ \Eprint
  {https://arxiv.org/abs/1203.6351} {arXiv:1203.6351 [hep-th]} \BibitemShut
  {NoStop}%
\bibitem [{\citenamefont {Assassi}\ \emph {et~al.}(2012)\citenamefont
  {Assassi}, \citenamefont {Baumann},\ and\ \citenamefont
  {Green}}]{Assassi:2012zq}%
  \BibitemOpen
  \bibfield  {author} {\bibinfo {author} {\bibfnamefont {V.}~\bibnamefont
  {Assassi}}, \bibinfo {author} {\bibfnamefont {D.}~\bibnamefont {Baumann}},\
  and\ \bibinfo {author} {\bibfnamefont {D.}~\bibnamefont {Green}},\ }\bibfield
   {title} {\bibinfo {title} {{On Soft Limits of Inflationary Correlation
  Functions}},\ }\href {https://doi.org/10.1088/1475-7516/2012/11/047}
  {\bibfield  {journal} {\bibinfo  {journal} {JCAP}\ }\textbf {\bibinfo
  {volume} {11}},\ \bibinfo {pages} {047}},\ \Eprint
  {https://arxiv.org/abs/1204.4207} {arXiv:1204.4207 [hep-th]} \BibitemShut
  {NoStop}%
\bibitem [{\citenamefont {Hinterbichler}\ \emph {et~al.}(2014)\citenamefont
  {Hinterbichler}, \citenamefont {Hui},\ and\ \citenamefont
  {Khoury}}]{Hinterbichler:2013dpa}%
  \BibitemOpen
  \bibfield  {author} {\bibinfo {author} {\bibfnamefont {K.}~\bibnamefont
  {Hinterbichler}}, \bibinfo {author} {\bibfnamefont {L.}~\bibnamefont {Hui}},\
  and\ \bibinfo {author} {\bibfnamefont {J.}~\bibnamefont {Khoury}},\
  }\bibfield  {title} {\bibinfo {title} {{An Infinite Set of Ward Identities
  for Adiabatic Modes in Cosmology}},\ }\href
  {https://doi.org/10.1088/1475-7516/2014/01/039} {\bibfield  {journal}
  {\bibinfo  {journal} {JCAP}\ }\textbf {\bibinfo {volume} {01}},\ \bibinfo
  {pages} {039}},\ \Eprint {https://arxiv.org/abs/1304.5527} {arXiv:1304.5527
  [hep-th]} \BibitemShut {NoStop}%
\bibitem [{\citenamefont {Goldberger}\ \emph {et~al.}(2013)\citenamefont
  {Goldberger}, \citenamefont {Hui},\ and\ \citenamefont
  {Nicolis}}]{Goldberger:2013rsa}%
  \BibitemOpen
  \bibfield  {author} {\bibinfo {author} {\bibfnamefont {W.~D.}\ \bibnamefont
  {Goldberger}}, \bibinfo {author} {\bibfnamefont {L.}~\bibnamefont {Hui}},\
  and\ \bibinfo {author} {\bibfnamefont {A.}~\bibnamefont {Nicolis}},\
  }\bibfield  {title} {\bibinfo {title} {{One-particle-irreducible consistency
  relations for cosmological perturbations}},\ }\href
  {https://doi.org/10.1103/PhysRevD.87.103520} {\bibfield  {journal} {\bibinfo
  {journal} {Phys. Rev. D}\ }\textbf {\bibinfo {volume} {87}},\ \bibinfo
  {pages} {103520} (\bibinfo {year} {2013})},\ \Eprint
  {https://arxiv.org/abs/1303.1193} {arXiv:1303.1193 [hep-th]} \BibitemShut
  {NoStop}%
\bibitem [{\citenamefont {Kundu}\ \emph {et~al.}(2015)\citenamefont {Kundu},
  \citenamefont {Shukla},\ and\ \citenamefont {Trivedi}}]{Kundu:2014gxa}%
  \BibitemOpen
  \bibfield  {author} {\bibinfo {author} {\bibfnamefont {N.}~\bibnamefont
  {Kundu}}, \bibinfo {author} {\bibfnamefont {A.}~\bibnamefont {Shukla}},\ and\
  \bibinfo {author} {\bibfnamefont {S.~P.}\ \bibnamefont {Trivedi}},\
  }\bibfield  {title} {\bibinfo {title} {{Constraints from Conformal Symmetry
  on the Three Point Scalar Correlator in Inflation}},\ }\href
  {https://doi.org/10.1007/JHEP04(2015)061} {\bibfield  {journal} {\bibinfo
  {journal} {JHEP}\ }\textbf {\bibinfo {volume} {04}},\ \bibinfo {pages}
  {061}},\ \Eprint {https://arxiv.org/abs/1410.2606} {arXiv:1410.2606 [hep-th]}
  \BibitemShut {NoStop}%
\bibitem [{\citenamefont {Kundu}\ \emph {et~al.}(2016)\citenamefont {Kundu},
  \citenamefont {Shukla},\ and\ \citenamefont {Trivedi}}]{Kundu:2015xta}%
  \BibitemOpen
  \bibfield  {author} {\bibinfo {author} {\bibfnamefont {N.}~\bibnamefont
  {Kundu}}, \bibinfo {author} {\bibfnamefont {A.}~\bibnamefont {Shukla}},\ and\
  \bibinfo {author} {\bibfnamefont {S.~P.}\ \bibnamefont {Trivedi}},\
  }\bibfield  {title} {\bibinfo {title} {{Ward Identities for Scale and Special
  Conformal Transformations in Inflation}},\ }\href
  {https://doi.org/10.1007/JHEP01(2016)046} {\bibfield  {journal} {\bibinfo
  {journal} {JHEP}\ }\textbf {\bibinfo {volume} {01}},\ \bibinfo {pages}
  {046}},\ \Eprint {https://arxiv.org/abs/1507.06017} {arXiv:1507.06017
  [hep-th]} \BibitemShut {NoStop}%
\bibitem [{\citenamefont {Shukla}\ \emph {et~al.}(2016)\citenamefont {Shukla},
  \citenamefont {Trivedi},\ and\ \citenamefont {Vishal}}]{Shukla:2016bnu}%
  \BibitemOpen
  \bibfield  {author} {\bibinfo {author} {\bibfnamefont {A.}~\bibnamefont
  {Shukla}}, \bibinfo {author} {\bibfnamefont {S.~P.}\ \bibnamefont
  {Trivedi}},\ and\ \bibinfo {author} {\bibfnamefont {V.}~\bibnamefont
  {Vishal}},\ }\bibfield  {title} {\bibinfo {title} {{Symmetry constraints in
  inflation, $\alpha$-vacua, and the three point function}},\ }\href
  {https://doi.org/10.1007/JHEP12(2016)102} {\bibfield  {journal} {\bibinfo
  {journal} {JHEP}\ }\textbf {\bibinfo {volume} {12}},\ \bibinfo {pages}
  {102}},\ \Eprint {https://arxiv.org/abs/1607.08636} {arXiv:1607.08636
  [hep-th]} \BibitemShut {NoStop}%
\bibitem [{\citenamefont {Gong}\ and\ \citenamefont
  {Seo}(2018)}]{Gong:2017wgx}%
  \BibitemOpen
  \bibfield  {author} {\bibinfo {author} {\bibfnamefont {J.-O.}\ \bibnamefont
  {Gong}}\ and\ \bibinfo {author} {\bibfnamefont {M.-S.}\ \bibnamefont {Seo}},\
  }\bibfield  {title} {\bibinfo {title} {{Consistency relations in multi-field
  inflation}},\ }\href {https://doi.org/10.1088/1475-7516/2018/02/008}
  {\bibfield  {journal} {\bibinfo  {journal} {JCAP}\ }\textbf {\bibinfo
  {volume} {02}},\ \bibinfo {pages} {008}},\ \Eprint
  {https://arxiv.org/abs/1707.08282} {arXiv:1707.08282 [hep-th]} \BibitemShut
  {NoStop}%
\bibitem [{\citenamefont {Pajer}\ and\ \citenamefont
  {Jazayeri}(2018)}]{Pajer:2017hmb}%
  \BibitemOpen
  \bibfield  {author} {\bibinfo {author} {\bibfnamefont {E.}~\bibnamefont
  {Pajer}}\ and\ \bibinfo {author} {\bibfnamefont {S.}~\bibnamefont
  {Jazayeri}},\ }\bibfield  {title} {\bibinfo {title} {{Systematics of
  Adiabatic Modes: Flat Universes}},\ }\href
  {https://doi.org/10.1088/1475-7516/2018/03/013} {\bibfield  {journal}
  {\bibinfo  {journal} {JCAP}\ }\textbf {\bibinfo {volume} {03}},\ \bibinfo
  {pages} {013}},\ \Eprint {https://arxiv.org/abs/1710.02177} {arXiv:1710.02177
  [astro-ph.CO]} \BibitemShut {NoStop}%
\bibitem [{\citenamefont {Bordin}\ \emph {et~al.}(2017)\citenamefont {Bordin},
  \citenamefont {Creminelli}, \citenamefont {Mirbabayi},\ and\ \citenamefont
  {Nore\~na}}]{Bordin:2017ozj}%
  \BibitemOpen
  \bibfield  {author} {\bibinfo {author} {\bibfnamefont {L.}~\bibnamefont
  {Bordin}}, \bibinfo {author} {\bibfnamefont {P.}~\bibnamefont {Creminelli}},
  \bibinfo {author} {\bibfnamefont {M.}~\bibnamefont {Mirbabayi}},\ and\
  \bibinfo {author} {\bibfnamefont {J.}~\bibnamefont {Nore\~na}},\ }\bibfield
  {title} {\bibinfo {title} {{Solid Consistency}},\ }\href
  {https://doi.org/10.1088/1475-7516/2017/03/004} {\bibfield  {journal}
  {\bibinfo  {journal} {JCAP}\ }\textbf {\bibinfo {volume} {03}},\ \bibinfo
  {pages} {004}},\ \Eprint {https://arxiv.org/abs/1701.04382} {arXiv:1701.04382
  [astro-ph.CO]} \BibitemShut {NoStop}%
\bibitem [{\citenamefont {Hui}\ \emph {et~al.}(2019)\citenamefont {Hui},
  \citenamefont {Joyce},\ and\ \citenamefont {Wong}}]{Hui:2018cag}%
  \BibitemOpen
  \bibfield  {author} {\bibinfo {author} {\bibfnamefont {L.}~\bibnamefont
  {Hui}}, \bibinfo {author} {\bibfnamefont {A.}~\bibnamefont {Joyce}},\ and\
  \bibinfo {author} {\bibfnamefont {S.~S.~C.}\ \bibnamefont {Wong}},\
  }\bibfield  {title} {\bibinfo {title} {{Inflationary soft theorems revisited:
  A generalized consistency relation}},\ }\href
  {https://doi.org/10.1088/1475-7516/2019/02/060} {\bibfield  {journal}
  {\bibinfo  {journal} {JCAP}\ }\textbf {\bibinfo {volume} {02}},\ \bibinfo
  {pages} {060}},\ \Eprint {https://arxiv.org/abs/1811.05951} {arXiv:1811.05951
  [hep-th]} \BibitemShut {NoStop}%
\bibitem [{\citenamefont {Jazayeri}\ \emph {et~al.}(2019)\citenamefont
  {Jazayeri}, \citenamefont {Pajer},\ and\ \citenamefont {van~der
  Woude}}]{Jazayeri:2019nbi}%
  \BibitemOpen
  \bibfield  {author} {\bibinfo {author} {\bibfnamefont {S.}~\bibnamefont
  {Jazayeri}}, \bibinfo {author} {\bibfnamefont {E.}~\bibnamefont {Pajer}},\
  and\ \bibinfo {author} {\bibfnamefont {D.}~\bibnamefont {van~der Woude}},\
  }\bibfield  {title} {\bibinfo {title} {{Solid Soft Theorems}},\ }\href
  {https://doi.org/10.1088/1475-7516/2019/06/011} {\bibfield  {journal}
  {\bibinfo  {journal} {JCAP}\ }\textbf {\bibinfo {volume} {06}},\ \bibinfo
  {pages} {011}},\ \Eprint {https://arxiv.org/abs/1902.09020} {arXiv:1902.09020
  [hep-th]} \BibitemShut {NoStop}%
\bibitem [{\citenamefont {Salopek}\ and\ \citenamefont
  {Bond}(1990)}]{Salopek:1990jq}%
  \BibitemOpen
  \bibfield  {author} {\bibinfo {author} {\bibfnamefont {D.~S.}\ \bibnamefont
  {Salopek}}\ and\ \bibinfo {author} {\bibfnamefont {J.~R.}\ \bibnamefont
  {Bond}},\ }\bibfield  {title} {\bibinfo {title} {{Nonlinear evolution of long
  wavelength metric fluctuations in inflationary models}},\ }\href
  {https://doi.org/10.1103/PhysRevD.42.3936} {\bibfield  {journal} {\bibinfo
  {journal} {Phys. Rev. D}\ }\textbf {\bibinfo {volume} {42}},\ \bibinfo
  {pages} {3936} (\bibinfo {year} {1990})}\BibitemShut {NoStop}%
\bibitem [{\citenamefont {Wands}\ \emph {et~al.}(2000)\citenamefont {Wands},
  \citenamefont {Malik}, \citenamefont {Lyth},\ and\ \citenamefont
  {Liddle}}]{Wands:2000dp}%
  \BibitemOpen
  \bibfield  {author} {\bibinfo {author} {\bibfnamefont {D.}~\bibnamefont
  {Wands}}, \bibinfo {author} {\bibfnamefont {K.~A.}\ \bibnamefont {Malik}},
  \bibinfo {author} {\bibfnamefont {D.~H.}\ \bibnamefont {Lyth}},\ and\
  \bibinfo {author} {\bibfnamefont {A.~R.}\ \bibnamefont {Liddle}},\ }\bibfield
   {title} {\bibinfo {title} {{A New approach to the evolution of cosmological
  perturbations on large scales}},\ }\href
  {https://doi.org/10.1103/PhysRevD.62.043527} {\bibfield  {journal} {\bibinfo
  {journal} {Phys. Rev. D}\ }\textbf {\bibinfo {volume} {62}},\ \bibinfo
  {pages} {043527} (\bibinfo {year} {2000})},\ \Eprint
  {https://arxiv.org/abs/astro-ph/0003278} {arXiv:astro-ph/0003278}
  \BibitemShut {NoStop}%
\bibitem [{\citenamefont {Weinberg}(2003)}]{Weinberg:2003sw}%
  \BibitemOpen
  \bibfield  {author} {\bibinfo {author} {\bibfnamefont {S.}~\bibnamefont
  {Weinberg}},\ }\bibfield  {title} {\bibinfo {title} {{Adiabatic modes in
  cosmology}},\ }\href {https://doi.org/10.1103/PhysRevD.67.123504} {\bibfield
  {journal} {\bibinfo  {journal} {Phys. Rev. D}\ }\textbf {\bibinfo {volume}
  {67}},\ \bibinfo {pages} {123504} (\bibinfo {year} {2003})},\ \Eprint
  {https://arxiv.org/abs/astro-ph/0302326} {arXiv:astro-ph/0302326}
  \BibitemShut {NoStop}%
\bibitem [{\citenamefont {Lyth}\ \emph {et~al.}(2005)\citenamefont {Lyth},
  \citenamefont {Malik},\ and\ \citenamefont {Sasaki}}]{Lyth:2004gb}%
  \BibitemOpen
  \bibfield  {author} {\bibinfo {author} {\bibfnamefont {D.~H.}\ \bibnamefont
  {Lyth}}, \bibinfo {author} {\bibfnamefont {K.~A.}\ \bibnamefont {Malik}},\
  and\ \bibinfo {author} {\bibfnamefont {M.}~\bibnamefont {Sasaki}},\
  }\bibfield  {title} {\bibinfo {title} {{A General proof of the conservation
  of the curvature perturbation}},\ }\href
  {https://doi.org/10.1088/1475-7516/2005/05/004} {\bibfield  {journal}
  {\bibinfo  {journal} {JCAP}\ }\textbf {\bibinfo {volume} {05}},\ \bibinfo
  {pages} {004}},\ \Eprint {https://arxiv.org/abs/astro-ph/0411220}
  {arXiv:astro-ph/0411220} \BibitemShut {NoStop}%
\bibitem [{\citenamefont {Weinberg}(2008)}]{Weinberg:2008zzc}%
  \BibitemOpen
  \bibfield  {author} {\bibinfo {author} {\bibfnamefont {S.}~\bibnamefont
  {Weinberg}},\ }\href@noop {} {\emph {\bibinfo {title} {{Cosmology}}}}\
  (\bibinfo {year} {2008})\BibitemShut {NoStop}%
\bibitem [{\citenamefont {Naruko}\ and\ \citenamefont
  {Sasaki}(2011)}]{Naruko:2011zk}%
  \BibitemOpen
  \bibfield  {author} {\bibinfo {author} {\bibfnamefont {A.}~\bibnamefont
  {Naruko}}\ and\ \bibinfo {author} {\bibfnamefont {M.}~\bibnamefont
  {Sasaki}},\ }\bibfield  {title} {\bibinfo {title} {{Conservation of the
  nonlinear curvature perturbation in generic single-field inflation}},\ }\href
  {https://doi.org/10.1088/0264-9381/28/7/072001} {\bibfield  {journal}
  {\bibinfo  {journal} {Class. Quant. Grav.}\ }\textbf {\bibinfo {volume}
  {28}},\ \bibinfo {pages} {072001} (\bibinfo {year} {2011})},\ \Eprint
  {https://arxiv.org/abs/1101.3180} {arXiv:1101.3180 [astro-ph.CO]}
  \BibitemShut {NoStop}%
\bibitem [{\citenamefont {Senatore}\ and\ \citenamefont
  {Zaldarriaga}(2013)}]{Senatore:2012ya}%
  \BibitemOpen
  \bibfield  {author} {\bibinfo {author} {\bibfnamefont {L.}~\bibnamefont
  {Senatore}}\ and\ \bibinfo {author} {\bibfnamefont {M.}~\bibnamefont
  {Zaldarriaga}},\ }\bibfield  {title} {\bibinfo {title} {{The constancy of
  $\zeta$ in single-clock Inflation at all loops}},\ }\href
  {https://doi.org/10.1007/JHEP09(2013)148} {\bibfield  {journal} {\bibinfo
  {journal} {JHEP}\ }\textbf {\bibinfo {volume} {09}},\ \bibinfo {pages}
  {148}},\ \Eprint {https://arxiv.org/abs/1210.6048} {arXiv:1210.6048 [hep-th]}
  \BibitemShut {NoStop}%
\bibitem [{\citenamefont {Starobinsky}(1982)}]{Starobinsky:1982ee}%
  \BibitemOpen
  \bibfield  {author} {\bibinfo {author} {\bibfnamefont {A.~A.}\ \bibnamefont
  {Starobinsky}},\ }\bibfield  {title} {\bibinfo {title} {{Dynamics of Phase
  Transition in the New Inflationary Universe Scenario and Generation of
  Perturbations}},\ }\href {https://doi.org/10.1016/0370-2693(82)90541-X}
  {\bibfield  {journal} {\bibinfo  {journal} {Phys. Lett. B}\ }\textbf
  {\bibinfo {volume} {117}},\ \bibinfo {pages} {175} (\bibinfo {year}
  {1982})}\BibitemShut {NoStop}%
\bibitem [{\citenamefont {Lyth}\ and\ \citenamefont
  {Seery}(2008)}]{Lyth:2006qz}%
  \BibitemOpen
  \bibfield  {author} {\bibinfo {author} {\bibfnamefont {D.~H.}\ \bibnamefont
  {Lyth}}\ and\ \bibinfo {author} {\bibfnamefont {D.}~\bibnamefont {Seery}},\
  }\bibfield  {title} {\bibinfo {title} {{Classicality of the primordial
  perturbations}},\ }\href {https://doi.org/10.1016/j.physletb.2008.03.010}
  {\bibfield  {journal} {\bibinfo  {journal} {Phys. Lett. B}\ }\textbf
  {\bibinfo {volume} {662}},\ \bibinfo {pages} {309} (\bibinfo {year}
  {2008})},\ \Eprint {https://arxiv.org/abs/astro-ph/0607647}
  {arXiv:astro-ph/0607647} \BibitemShut {NoStop}%
\bibitem [{\citenamefont {Baldauf}\ \emph {et~al.}(2011)\citenamefont
  {Baldauf}, \citenamefont {Seljak}, \citenamefont {Senatore},\ and\
  \citenamefont {Zaldarriaga}}]{Baldauf:2011bh}%
  \BibitemOpen
  \bibfield  {author} {\bibinfo {author} {\bibfnamefont {T.}~\bibnamefont
  {Baldauf}}, \bibinfo {author} {\bibfnamefont {U.}~\bibnamefont {Seljak}},
  \bibinfo {author} {\bibfnamefont {L.}~\bibnamefont {Senatore}},\ and\
  \bibinfo {author} {\bibfnamefont {M.}~\bibnamefont {Zaldarriaga}},\
  }\bibfield  {title} {\bibinfo {title} {{Galaxy Bias and non-Linear Structure
  Formation in General Relativity}},\ }\href
  {https://doi.org/10.1088/1475-7516/2011/10/031} {\bibfield  {journal}
  {\bibinfo  {journal} {JCAP}\ }\textbf {\bibinfo {volume} {10}},\ \bibinfo
  {pages} {031}},\ \Eprint {https://arxiv.org/abs/1106.5507} {arXiv:1106.5507
  [astro-ph.CO]} \BibitemShut {NoStop}%
\bibitem [{\citenamefont {Tanaka}\ and\ \citenamefont
  {Urakawa}(2011)}]{Tanaka:2011aj}%
  \BibitemOpen
  \bibfield  {author} {\bibinfo {author} {\bibfnamefont {T.}~\bibnamefont
  {Tanaka}}\ and\ \bibinfo {author} {\bibfnamefont {Y.}~\bibnamefont
  {Urakawa}},\ }\bibfield  {title} {\bibinfo {title} {{Dominance of gauge
  artifact in the consistency relation for the primordial bispectrum}},\ }\href
  {https://doi.org/10.1088/1475-7516/2011/05/014} {\bibfield  {journal}
  {\bibinfo  {journal} {JCAP}\ }\textbf {\bibinfo {volume} {05}},\ \bibinfo
  {pages} {014}},\ \Eprint {https://arxiv.org/abs/1103.1251} {arXiv:1103.1251
  [astro-ph.CO]} \BibitemShut {NoStop}%
\bibitem [{\citenamefont {Pimentel}\ \emph {et~al.}(2012)\citenamefont
  {Pimentel}, \citenamefont {Senatore},\ and\ \citenamefont
  {Zaldarriaga}}]{Pimentel:2012tw}%
  \BibitemOpen
  \bibfield  {author} {\bibinfo {author} {\bibfnamefont {G.~L.}\ \bibnamefont
  {Pimentel}}, \bibinfo {author} {\bibfnamefont {L.}~\bibnamefont {Senatore}},\
  and\ \bibinfo {author} {\bibfnamefont {M.}~\bibnamefont {Zaldarriaga}},\
  }\bibfield  {title} {\bibinfo {title} {{On Loops in Inflation III: Time
  Independence of zeta in Single Clock Inflation}},\ }\href
  {https://doi.org/10.1007/JHEP07(2012)166} {\bibfield  {journal} {\bibinfo
  {journal} {JHEP}\ }\textbf {\bibinfo {volume} {07}},\ \bibinfo {pages}
  {166}},\ \Eprint {https://arxiv.org/abs/1203.6651} {arXiv:1203.6651 [hep-th]}
  \BibitemShut {NoStop}%
\bibitem [{\citenamefont {Pajer}\ \emph {et~al.}(2013)\citenamefont {Pajer},
  \citenamefont {Schmidt},\ and\ \citenamefont {Zaldarriaga}}]{Pajer:2013ana}%
  \BibitemOpen
  \bibfield  {author} {\bibinfo {author} {\bibfnamefont {E.}~\bibnamefont
  {Pajer}}, \bibinfo {author} {\bibfnamefont {F.}~\bibnamefont {Schmidt}},\
  and\ \bibinfo {author} {\bibfnamefont {M.}~\bibnamefont {Zaldarriaga}},\
  }\bibfield  {title} {\bibinfo {title} {{The Observed Squeezed Limit of
  Cosmological Three-Point Functions}},\ }\href
  {https://doi.org/10.1103/PhysRevD.88.083502} {\bibfield  {journal} {\bibinfo
  {journal} {Phys. Rev. D}\ }\textbf {\bibinfo {volume} {88}},\ \bibinfo
  {pages} {083502} (\bibinfo {year} {2013})},\ \Eprint
  {https://arxiv.org/abs/1305.0824} {arXiv:1305.0824 [astro-ph.CO]}
  \BibitemShut {NoStop}%
\bibitem [{\citenamefont {de~Putter}\ \emph {et~al.}(2015)\citenamefont
  {de~Putter}, \citenamefont {Dor\'e},\ and\ \citenamefont
  {Green}}]{dePutter:2015vga}%
  \BibitemOpen
  \bibfield  {author} {\bibinfo {author} {\bibfnamefont {R.}~\bibnamefont
  {de~Putter}}, \bibinfo {author} {\bibfnamefont {O.}~\bibnamefont {Dor\'e}},\
  and\ \bibinfo {author} {\bibfnamefont {D.}~\bibnamefont {Green}},\ }\bibfield
   {title} {\bibinfo {title} {{Is There Scale-Dependent Bias in Single-Field
  Inflation?}},\ }\href {https://doi.org/10.1088/1475-7516/2015/10/024}
  {\bibfield  {journal} {\bibinfo  {journal} {JCAP}\ }\textbf {\bibinfo
  {volume} {10}},\ \bibinfo {pages} {024}},\ \Eprint
  {https://arxiv.org/abs/1504.05935} {arXiv:1504.05935 [astro-ph.CO]}
  \BibitemShut {NoStop}%
\bibitem [{\citenamefont {Dai}\ \emph {et~al.}(2015{\natexlab{a}})\citenamefont
  {Dai}, \citenamefont {Pajer},\ and\ \citenamefont {Schmidt}}]{Dai:2015jaa}%
  \BibitemOpen
  \bibfield  {author} {\bibinfo {author} {\bibfnamefont {L.}~\bibnamefont
  {Dai}}, \bibinfo {author} {\bibfnamefont {E.}~\bibnamefont {Pajer}},\ and\
  \bibinfo {author} {\bibfnamefont {F.}~\bibnamefont {Schmidt}},\ }\bibfield
  {title} {\bibinfo {title} {{On Separate Universes}},\ }\href
  {https://doi.org/10.1088/1475-7516/2015/10/059} {\bibfield  {journal}
  {\bibinfo  {journal} {JCAP}\ }\textbf {\bibinfo {volume} {10}},\ \bibinfo
  {pages} {059}},\ \Eprint {https://arxiv.org/abs/1504.00351} {arXiv:1504.00351
  [astro-ph.CO]} \BibitemShut {NoStop}%
\bibitem [{\citenamefont {Dai}\ \emph {et~al.}(2015{\natexlab{b}})\citenamefont
  {Dai}, \citenamefont {Pajer},\ and\ \citenamefont {Schmidt}}]{Dai:2015rda}%
  \BibitemOpen
  \bibfield  {author} {\bibinfo {author} {\bibfnamefont {L.}~\bibnamefont
  {Dai}}, \bibinfo {author} {\bibfnamefont {E.}~\bibnamefont {Pajer}},\ and\
  \bibinfo {author} {\bibfnamefont {F.}~\bibnamefont {Schmidt}},\ }\bibfield
  {title} {\bibinfo {title} {{Conformal Fermi Coordinates}},\ }\href
  {https://doi.org/10.1088/1475-7516/2015/11/043} {\bibfield  {journal}
  {\bibinfo  {journal} {JCAP}\ }\textbf {\bibinfo {volume} {11}},\ \bibinfo
  {pages} {043}},\ \Eprint {https://arxiv.org/abs/1502.02011} {arXiv:1502.02011
  [gr-qc]} \BibitemShut {NoStop}%
\bibitem [{\citenamefont {Cabass}\ \emph {et~al.}(2017)\citenamefont {Cabass},
  \citenamefont {Pajer},\ and\ \citenamefont {Schmidt}}]{Cabass:2016cgp}%
  \BibitemOpen
  \bibfield  {author} {\bibinfo {author} {\bibfnamefont {G.}~\bibnamefont
  {Cabass}}, \bibinfo {author} {\bibfnamefont {E.}~\bibnamefont {Pajer}},\ and\
  \bibinfo {author} {\bibfnamefont {F.}~\bibnamefont {Schmidt}},\ }\bibfield
  {title} {\bibinfo {title} {{How Gaussian can our Universe be?}},\ }\href
  {https://doi.org/10.1088/1475-7516/2017/01/003} {\bibfield  {journal}
  {\bibinfo  {journal} {JCAP}\ }\textbf {\bibinfo {volume} {01}},\ \bibinfo
  {pages} {003}},\ \Eprint {https://arxiv.org/abs/1612.00033} {arXiv:1612.00033
  [hep-th]} \BibitemShut {NoStop}%
\bibitem [{\citenamefont {Bravo}\ \emph
  {et~al.}(2018{\natexlab{b}})\citenamefont {Bravo}, \citenamefont {Mooij},
  \citenamefont {Palma},\ and\ \citenamefont {Pradenas}}]{Bravo:2017gct}%
  \BibitemOpen
  \bibfield  {author} {\bibinfo {author} {\bibfnamefont {R.}~\bibnamefont
  {Bravo}}, \bibinfo {author} {\bibfnamefont {S.}~\bibnamefont {Mooij}},
  \bibinfo {author} {\bibfnamefont {G.~A.}\ \bibnamefont {Palma}},\ and\
  \bibinfo {author} {\bibfnamefont {B.}~\bibnamefont {Pradenas}},\ }\bibfield
  {title} {\bibinfo {title} {{Vanishing of local non-Gaussianity in canonical
  single field inflation}},\ }\href
  {https://doi.org/10.1088/1475-7516/2018/05/025} {\bibfield  {journal}
  {\bibinfo  {journal} {JCAP}\ }\textbf {\bibinfo {volume} {05}},\ \bibinfo
  {pages} {025}},\ \Eprint {https://arxiv.org/abs/1711.05290} {arXiv:1711.05290
  [astro-ph.CO]} \BibitemShut {NoStop}%
\bibitem [{\citenamefont {Cabass}\ \emph {et~al.}(2018)\citenamefont {Cabass},
  \citenamefont {Pajer},\ and\ \citenamefont {Schmidt}}]{Cabass:2018roz}%
  \BibitemOpen
  \bibfield  {author} {\bibinfo {author} {\bibfnamefont {G.}~\bibnamefont
  {Cabass}}, \bibinfo {author} {\bibfnamefont {E.}~\bibnamefont {Pajer}},\ and\
  \bibinfo {author} {\bibfnamefont {F.}~\bibnamefont {Schmidt}},\ }\bibfield
  {title} {\bibinfo {title} {{Imprints of Oscillatory Bispectra on Galaxy
  Clustering}},\ }\href {https://doi.org/10.1088/1475-7516/2018/09/003}
  {\bibfield  {journal} {\bibinfo  {journal} {JCAP}\ }\textbf {\bibinfo
  {volume} {09}},\ \bibinfo {pages} {003}},\ \Eprint
  {https://arxiv.org/abs/1804.07295} {arXiv:1804.07295 [astro-ph.CO]}
  \BibitemShut {NoStop}%
\bibitem [{\citenamefont {Umeh}\ \emph {et~al.}(2019)\citenamefont {Umeh},
  \citenamefont {Koyama}, \citenamefont {Maartens}, \citenamefont {Schmidt},\
  and\ \citenamefont {Clarkson}}]{Umeh:2019qyd}%
  \BibitemOpen
  \bibfield  {author} {\bibinfo {author} {\bibfnamefont {O.}~\bibnamefont
  {Umeh}}, \bibinfo {author} {\bibfnamefont {K.}~\bibnamefont {Koyama}},
  \bibinfo {author} {\bibfnamefont {R.}~\bibnamefont {Maartens}}, \bibinfo
  {author} {\bibfnamefont {F.}~\bibnamefont {Schmidt}},\ and\ \bibinfo {author}
  {\bibfnamefont {C.}~\bibnamefont {Clarkson}},\ }\bibfield  {title} {\bibinfo
  {title} {{General relativistic effects in the galaxy bias at second order}},\
  }\href {https://doi.org/10.1088/1475-7516/2019/05/020} {\bibfield  {journal}
  {\bibinfo  {journal} {JCAP}\ }\textbf {\bibinfo {volume} {05}},\ \bibinfo
  {pages} {020}},\ \Eprint {https://arxiv.org/abs/1901.07460} {arXiv:1901.07460
  [astro-ph.CO]} \BibitemShut {NoStop}%
\bibitem [{\citenamefont {Mitsou}\ \emph {et~al.}(2021)\citenamefont {Mitsou},
  \citenamefont {Fanizza}, \citenamefont {Grimm},\ and\ \citenamefont
  {Yoo}}]{Mitsou:2020czr}%
  \BibitemOpen
  \bibfield  {author} {\bibinfo {author} {\bibfnamefont {E.}~\bibnamefont
  {Mitsou}}, \bibinfo {author} {\bibfnamefont {G.}~\bibnamefont {Fanizza}},
  \bibinfo {author} {\bibfnamefont {N.}~\bibnamefont {Grimm}},\ and\ \bibinfo
  {author} {\bibfnamefont {J.}~\bibnamefont {Yoo}},\ }\bibfield  {title}
  {\bibinfo {title} {{Cutting out the cosmological middle man: General
  Relativity in the light-cone coordinates}},\ }\href
  {https://doi.org/10.1088/1361-6382/abd681} {\bibfield  {journal} {\bibinfo
  {journal} {Class. Quant. Grav.}\ }\textbf {\bibinfo {volume} {38}},\ \bibinfo
  {pages} {055011} (\bibinfo {year} {2021})},\ \Eprint
  {https://arxiv.org/abs/2009.14687} {arXiv:2009.14687 [gr-qc]} \BibitemShut
  {NoStop}%
\bibitem [{\citenamefont {Bardeen}(1980)}]{Bardeen:1980kt}%
  \BibitemOpen
  \bibfield  {author} {\bibinfo {author} {\bibfnamefont {J.~M.}\ \bibnamefont
  {Bardeen}},\ }\bibfield  {title} {\bibinfo {title} {{Gauge Invariant
  Cosmological Perturbations}},\ }\href
  {https://doi.org/10.1103/PhysRevD.22.1882} {\bibfield  {journal} {\bibinfo
  {journal} {Phys. Rev. D}\ }\textbf {\bibinfo {volume} {22}},\ \bibinfo
  {pages} {1882} (\bibinfo {year} {1980})}\BibitemShut {NoStop}%
\bibitem [{\citenamefont {Endlich}\ \emph {et~al.}(2013)\citenamefont
  {Endlich}, \citenamefont {Nicolis},\ and\ \citenamefont
  {Wang}}]{Endlich:2012pz}%
  \BibitemOpen
  \bibfield  {author} {\bibinfo {author} {\bibfnamefont {S.}~\bibnamefont
  {Endlich}}, \bibinfo {author} {\bibfnamefont {A.}~\bibnamefont {Nicolis}},\
  and\ \bibinfo {author} {\bibfnamefont {J.}~\bibnamefont {Wang}},\ }\bibfield
  {title} {\bibinfo {title} {{Solid Inflation}},\ }\href
  {https://doi.org/10.1088/1475-7516/2013/10/011} {\bibfield  {journal}
  {\bibinfo  {journal} {JCAP}\ }\textbf {\bibinfo {volume} {10}},\ \bibinfo
  {pages} {011}},\ \Eprint {https://arxiv.org/abs/1210.0569} {arXiv:1210.0569
  [hep-th]} \BibitemShut {NoStop}%
\bibitem [{\citenamefont {Endlich}\ \emph {et~al.}(2014)\citenamefont
  {Endlich}, \citenamefont {Horn}, \citenamefont {Nicolis},\ and\ \citenamefont
  {Wang}}]{Endlich:2013jia}%
  \BibitemOpen
  \bibfield  {author} {\bibinfo {author} {\bibfnamefont {S.}~\bibnamefont
  {Endlich}}, \bibinfo {author} {\bibfnamefont {B.}~\bibnamefont {Horn}},
  \bibinfo {author} {\bibfnamefont {A.}~\bibnamefont {Nicolis}},\ and\ \bibinfo
  {author} {\bibfnamefont {J.}~\bibnamefont {Wang}},\ }\bibfield  {title}
  {\bibinfo {title} {{Squeezed limit of the solid inflation three-point
  function}},\ }\href {https://doi.org/10.1103/PhysRevD.90.063506} {\bibfield
  {journal} {\bibinfo  {journal} {Phys. Rev. D}\ }\textbf {\bibinfo {volume}
  {90}},\ \bibinfo {pages} {063506} (\bibinfo {year} {2014})},\ \Eprint
  {https://arxiv.org/abs/1307.8114} {arXiv:1307.8114 [hep-th]} \BibitemShut
  {NoStop}%
\bibitem [{\citenamefont {Chen}\ \emph {et~al.}(2007)\citenamefont {Chen},
  \citenamefont {Huang}, \citenamefont {Kachru},\ and\ \citenamefont
  {Shiu}}]{Chen:2006nt}%
  \BibitemOpen
  \bibfield  {author} {\bibinfo {author} {\bibfnamefont {X.}~\bibnamefont
  {Chen}}, \bibinfo {author} {\bibfnamefont {M.-x.}\ \bibnamefont {Huang}},
  \bibinfo {author} {\bibfnamefont {S.}~\bibnamefont {Kachru}},\ and\ \bibinfo
  {author} {\bibfnamefont {G.}~\bibnamefont {Shiu}},\ }\bibfield  {title}
  {\bibinfo {title} {{Observational signatures and non-Gaussianities of general
  single field inflation}},\ }\href
  {https://doi.org/10.1088/1475-7516/2007/01/002} {\bibfield  {journal}
  {\bibinfo  {journal} {JCAP}\ }\textbf {\bibinfo {volume} {01}},\ \bibinfo
  {pages} {002}},\ \Eprint {https://arxiv.org/abs/hep-th/0605045}
  {arXiv:hep-th/0605045} \BibitemShut {NoStop}%
\bibitem [{\citenamefont {Bartolo}\ \emph {et~al.}(2010)\citenamefont
  {Bartolo}, \citenamefont {Fasiello}, \citenamefont {Matarrese},\ and\
  \citenamefont {Riotto}}]{Bartolo:2010im}%
  \BibitemOpen
  \bibfield  {author} {\bibinfo {author} {\bibfnamefont {N.}~\bibnamefont
  {Bartolo}}, \bibinfo {author} {\bibfnamefont {M.}~\bibnamefont {Fasiello}},
  \bibinfo {author} {\bibfnamefont {S.}~\bibnamefont {Matarrese}},\ and\
  \bibinfo {author} {\bibfnamefont {A.}~\bibnamefont {Riotto}},\ }\bibfield
  {title} {\bibinfo {title} {{Tilt and Running of Cosmological Observables in
  Generalized Single-Field Inflation}},\ }\href
  {https://doi.org/10.1088/1475-7516/2010/12/026} {\bibfield  {journal}
  {\bibinfo  {journal} {JCAP}\ }\textbf {\bibinfo {volume} {12}},\ \bibinfo
  {pages} {026}},\ \Eprint {https://arxiv.org/abs/1010.3993} {arXiv:1010.3993
  [astro-ph.CO]} \BibitemShut {NoStop}%
\bibitem [{\citenamefont {Pajer}(2021)}]{Pajer:2020wxk}%
  \BibitemOpen
  \bibfield  {author} {\bibinfo {author} {\bibfnamefont {E.}~\bibnamefont
  {Pajer}},\ }\bibfield  {title} {\bibinfo {title} {{Building a Boostless
  Bootstrap for the Bispectrum}},\ }\href
  {https://doi.org/10.1088/1475-7516/2021/01/023} {\bibfield  {journal}
  {\bibinfo  {journal} {JCAP}\ }\textbf {\bibinfo {volume} {01}},\ \bibinfo
  {pages} {023}},\ \Eprint {https://arxiv.org/abs/2010.12818} {arXiv:2010.12818
  [hep-th]} \BibitemShut {NoStop}%
\bibitem [{\citenamefont {Maldacena}\ and\ \citenamefont
  {Pimentel}(2011)}]{Maldacena:2011nz}%
  \BibitemOpen
  \bibfield  {author} {\bibinfo {author} {\bibfnamefont {J.~M.}\ \bibnamefont
  {Maldacena}}\ and\ \bibinfo {author} {\bibfnamefont {G.~L.}\ \bibnamefont
  {Pimentel}},\ }\bibfield  {title} {\bibinfo {title} {{On graviton
  non-Gaussianities during inflation}},\ }\href
  {https://doi.org/10.1007/JHEP09(2011)045} {\bibfield  {journal} {\bibinfo
  {journal} {JHEP}\ }\textbf {\bibinfo {volume} {09}},\ \bibinfo {pages}
  {045}},\ \Eprint {https://arxiv.org/abs/1104.2846} {arXiv:1104.2846 [hep-th]}
  \BibitemShut {NoStop}%
\end{thebibliography}%
